\documentclass{aa}

\usepackage[T1]{fontenc}
\usepackage[utf8]{inputenc}

\usepackage{graphicx}
\usepackage{natbib}
\usepackage{ulem}
\usepackage{textcomp}
\usepackage{gensymb}
\usepackage{longtable}
\usepackage{threeparttable}
\usepackage{multicol}
\usepackage{multirow}
\usepackage{textgreek}
\usepackage{float}
\usepackage[Symbol]{upgreek}
\usepackage{amsmath}
\usepackage{etoolbox}
\usepackage{txfonts}
\usepackage{url}
\usepackage{xcolor}

\usepackage[most]{tcolorbox}
\usepackage{hyperref}
\usepackage{xcolor}
\hypersetup{
  colorlinks   = true, 
  urlcolor     = black, 
  linkcolor    = blue, 
  citecolor   = blue 
}
\usepackage[all]{hypcap} 

\begin{document}

	\title{New mysteries and challenges from the Toothbrush relic: \\
	 wideband observations from 550\,MHz to 8\,GHz}

	\titlerunning{Deep wideband observations of 1RXS\,J0603.3+4214 in the frequency range 550\,MHz to 8\,GHz}
	\authorrunning{Rajpurohit et al.}

\author{K. Rajpurohit\inst{1,2}, M. Hoeft \inst{3}, F. Vazza\inst{1,2,4}, L. Rudnick\inst{5}, R. J. van Weeren\inst{6}, D. Wittor\inst{1,2,4}, A. Drabent\inst{3}\\ M. Brienza\inst{1,2}, , E. Bonnassieux\inst{1}, N. Locatelli\inst{1}, R. Kale\inst{7}, and C. Dumba\inst{8} }
\institute{Dipartimento di Fisica e Astronomia, Universit\'at di Bologna, via P. Gobetti 93/2, I-40129, Bologna, Italy\\
 {\email{kamlesh.rajpurohit@unibo.it}}
\and
INAF-Istituto di Radio Astronomia, Via Gobetti 101, Bologna, Italy
\and
Th\"{u}ringer Landessternwarte (TLS), Sternwarte 5, 07778 Tautenburg, Germany
\and
Hamburger Sternwarte, Universit\"at Hamburg, Gojenbergsweg 112, D-21029, Hamburg, Germany
\and
Minnesota Institute for Astrophysics, University of Minnesota, 116 Church St. S.E., Minneapolis, MN 55455, USA
\and 
Leiden Observatory, Leiden University, P.O. Box 9513, NL-2300 RA Leiden, The Netherlands
\and 
National Centre for Radio Astrophysics, Tata Institute of Fundamental Research, P. B. No. 3, Ganeshkhind, Pune 411007, India
\and 
Mbarara University of Science and Technology, Mbarara, Uganda
}	
        \abstract
   {Radio relics are diffuse extended synchrotron sources that originate from shock fronts induced by galaxy cluster mergers. The actual particle acceleration mechanism at the shock fronts is still under debate. The galaxy cluster 1RXS\,J0603.3+4214 hosts one of the most intriguing examples of radio relics, known as the Toothbrush.}
   { To understand the mechanism(s) that accelerate(s) relativistic particles in the intracluster medium (ICM), we investigate the spectral properties of large scale diffuse extended sources in the merging galaxy cluster 1RXS\,J0603.3+4214. }
   {We present new wideband radio continuum observations made with uGMRT and VLA. Our new observations, in combination with previously published data, allowed us to carry out a detailed high spatial resolution spectral and curvature analysis of the known diffuse radio emission sources, using a wide range of frequencies.}
   { The integrated spectrum of the Toothbrush follows closely a power law over almost two orders of magnitudes in frequency, with a spectral index of $-1.16\pm0.02$. We do not find any evidence of spectral steepening below 8\,GHz.  The subregions of the Toothbrush also exhibit near-perfect power laws and an identical spectral slopes, suggesting that observed spectral index is rather set by the distribution of Mach numbers which may have a similar shape at different parts of the shock front. Indeed, numerical simulations show an intriguing similar spectral index, indicating that the radio spectrum is dominated by the average over the inhomogeneities within the shock, with most of the emission coming from the tail of the Mach number distribution. In contrast to the Toothbrush, the spectrum of the fainter relics show a high-frequency steepening. Moreover, also the integrated spectrum of the halo follows a power law from 150\,MHz to 3\,GHz with a spectral index of  $-1.16\pm0.04$.  We do not find any evidence for spectral curvature, not even in subareas of the halo. This suggest a homogeneous acceleration throughout the cluster volume.  
   Between the "brush" region of the Toothbrush and the halo, the color-color analysis revealed emission that was consistent with an overlap between the two different spectral regions.
   }
   { None of the relic structures, the Toothbrush as a whole or its subregions or the other two fainter relics, show spectral shapes consistent with a single injection of relativistic electrons, such as at a shock, followed by synchrotron aging in a relatively homogeneous environment.
   Inhomogeneities in some combination of Mach number, magnetic field strengths and projection effects dominate the observed spectral shapes.}


   \maketitle
%


\section{Introduction} \label{sec:intro}

Radio observations of a fraction of galaxy clusters reveal the presence of spectacular large scale diffuse synchrotron sources. These diffuse sources are known as radio relics and radio halos \citep[see][for a recent review]{vanWeeren2019}. They show a typical synchrotron spectrum\footnote{$S(\nu)\propto\nu^{{\alpha}}$ with spectral index $\alpha$}. In the hierarchical model of structure formation, galaxy clusters form through a sequence of mergers with smaller substructures. During cluster mergers, shock waves and turbulence are produced within the intracluster medium (ICM). Merger driven shocks and turbulence are expected to be efficient accelerators of cosmic rays (CRs) to relativistic energies and may amplify cluster's magnetic fields. 

Radio halos are typically unpolarized sources located in the center of galaxy clusters and have an extent of $\sim$\,Mpc. They roughly follow the X-ray emission from the ICM. The currently favored scenario for the formation of radio halos involves the re-acceleration of cosmic-ray electrons (CRe) via magneto-hydrodynamical turbulence \citep[e.g.][]{Brunetti2001,Petrosian2001,Brunetti2014}.

\setlength{\tabcolsep}{12pt}
\begin{table*}[!htbp]
\caption{Observations overview }
\centering
\begin{threeparttable} 
\begin{tabular}{ c | c  c | c | c   }
\hline \hline
\multirow{1}{*}{} & \multicolumn{2}{c|}{VLA S-band}  & \multirow{1}{*}{VLA C-band} &  \multirow{1}{*}{uGMRT}\\
 \cline{2-3} \cline{4-5} 
  &B-configuration& C-configuration & C-configuration& band\,4\\
  \hline  
Observation date& September 9, 2017 & November 29, 2018 & November 18, 2018 & September 14, 2018  \\
Frequency range&2-4\,GHz& 2-4\,GHz & 4-8\,GHz &550-750\,MHz\\ 
Channel width & 2\,MHz & 2\,MHz & 27\,MHz &49 kHz\\ 
No. of IF &16 & 16 & 36 &1\\ 
No. of channels per IF &64 & 64& 64 &4096\\ 
Integration time& 3\,s & 5\,s& 2\,s & 8\,s  \\
On-source time &3\,hrs &5\,hrs &5\,hrs &8\,hrs \\
\hline 
\end{tabular}
\begin{tablenotes}[flushleft]
\footnotesize
\item{\textbf{Notes.}} Full Stokes polarization information was recorded for the VLA  and the uGMRT observations.
\end{tablenotes}
\end{threeparttable} 
\label{Tabel:Tabel1}   
\end{table*}

Radio relics are Mpc-sized elongated radio sources typically found in the periphery of merging galaxy clusters \citep[see][for a review]{vanWeeren2019}. They are usually strongly polarized \citep{Bonafede2009,Bonafede2012,vanweeren2011,vanWeeren2012a,Kale2012,Owen2014}. Relics are believed to originate from shock fronts induced by galaxy cluster mergers. This connection has been confirmed by sometimes finding shock fronts in the X-ray surface brightness distribution located where radio relics have been found \citep[see e.g.,][]{Sarazin2013,Ogrean2013,Shimwell2015,vanWeeren2016,Botteon2016,Tholken2018,Gennaro2019}. The sizes of relics and their separations from the cluster center vary significantly, as well as their degree of association with shocks detected via X-ray observations, which makes it difficult to obtain a self-consistent theoretical model of their origin.

Relic emission is believed to be produced by diffusive shock acceleration (DSA) at the merger shocks \citep[e.g.][for a recent review]{2019SSRv..215...14B}. Theoretical models predicting the observed radio power as a function of shock parameters have shown that the physical connection between shocks and radio emission is physically viable \citep[e.g.][]{Ensslin1998,Hoeft2007,ka12}. Numerical simulations have  also been able to produce radio relics that reasonably resemble the observed ones in shape, total power and spectral index. \citep[][]{ho08,2009MNRAS.393.1073B,sk11,2017MNRAS.470..240N}. 

However, several open problems arise if DSA is used to make quantitative predictions, pointing to limitations of the simplest models. There are three major difficulties: (1) the Mach numbers derived from X-ray observations are often significantly lower than those derived from radio observations; (2) DSA operating at shock fronts with Mach numbers $\mathcal{M}\,\leq4$  requires an unrealistically high CRe acceleration efficiency in order to explain the relic luminosities \citep[e.g.][]{va14relics,va15relics}; (3) there is evidence that the spectral index of some relics shows steepening at higher frequencies \citep{Stroe2016,Malu2016} or a flat integrated spectrum at low frequencies \citep{trasatti2015,Kierdorf2016}, which is incompatible with the simplest models using DSA in a steady state.

An alternative mechanism is that mildly relativistic fossil electrons, perhaps from old radio galaxies, are reaccelerated in the shock \citep{Kang2016a}. In this scenario, weaker shocks are able to accelerate the population of aged CRe. A volume filling population of fossil ($\geq 10^8 \rm ~yr$) electrons which gets re-accelerated by shocks would alleviate the efficiency problem \citep[e.g.][]{2013MNRAS.435.1061P,ka12}. This mechanism also predicts spectral steepening at high frequencies.

In the past decade, numerous studies have been dedicated to characterize the integrated spectra of relics \citep{Itahana2015,Stroe2016,Kierdorf2016}. The major question has been: Is there any steepening of the spectrum at high frequencies? These studies have provided different answers, owing to the differences in the covered frequency range, uv-coverage, and the type of observations (interferometric or single dish). The question of whether or not such a steepening exists is extremely important since it sheds light on the mechanisms which generate radio relics.

\setlength{\tabcolsep}{8.0pt}   
\begin{table*}[!htbp]
\caption{Image properties }
\centering
\label{Table 2}
 \begin{threeparttable} 
\begin{tabular}{c c c c c c c r}
\hline\hline
Band & Configuration  & Name &Restoring Beam & Weighting & uv-cut & uv-taper & RMS noise\\ 
&&&&&&&$\upmu\,\rm Jy\,beam^{-1}$\\
\hline
   & C&IM1&$2\farcs7 \times 2\farcs6$&Briggs&none&none&3.2\\
VLA C-band & C&IM2&$5\farcs5 \times 5\farcs5$&Uniform&$ \geq\rm0.4\,k\uplambda$&5\arcsec& 5.6\\
(4-8\,GHz)&C&IM3&$8\farcs0 \times 8\farcs0$&Uniform&$ \geq\rm0.4\,k\uplambda$&9\arcsec&7.1\\
\hline   
 &BC&IM4&$2\farcs0 \times 1\farcs8$&Briggs&none&none&3.5\\
&  BC&IM5&$5\farcs0 \times 5\farcs0$&Briggs&none&6\arcsec& 4.8\\
&  BC&IM6&$5\farcs5 \times 5\farcs5$&Uniform&$ \geq\rm0.4\,k\uplambda$&6\arcsec&5.6 \\
VLA S-band&BC&IM7&$8\farcs0 \times 8\farcs0$&Uniform&$ \geq\rm0.4\,k\uplambda$&9\arcsec&6.4\\
(2-4\,GHz) &  BC&IM8&$10\arcsec \times 10\arcsec$&Briggs& none&11\arcsec&7.8\\
 &  BC&IM9&$15\arcsec \times 15\arcsec$&Briggs&none&16\arcsec&8.3\\
  & BC&IM10&$15\farcs7\times15\farcs7$& Unifrom&$\geq\rm0.4\,k\uplambda$&16\arcsec&9.1\\
\hline
& ABCD&IM11&$5\farcs5 \times 5\farcs5$&Uniform&$ \geq\rm0.4\,k\uplambda$&6\arcsec&7.1\\
VLA L-band$^{\dagger}$&  ABCD&IM12&$8\farcs0 \times 8\farcs0$&Uniform&$ \geq\rm0.4\,k\uplambda$&9\arcsec&9.3\\
(1-2\,GHz) &  ABCD&IM13&$15\farcs7\times15\farcs7$&Uniform&$ \geq\rm0.4\,k\uplambda$&16\arcsec&16.2\\
\hline
  &  band 4&IM14&$5\farcs5 \times 4\farcs8$&Briggs&none&none&7.7\\
  &  band 4&IM15&$5\farcs5 \times 5\farcs5$&Uniform&$\geq\rm0.4\,k\uplambda$&5\arcsec&16\\
uGMRT& band 4&IM16&$8\farcs0 \times 8\farcs0$&Uniform&$ \geq\rm0.4\,k\uplambda$&8\arcsec&18.2\\
 (550-750 MHz)&  band 4&IM17&$15\arcsec \times 15\arcsec$&Briggs&none&15\arcsec&41.1\\
 &  band 4&IM18&$15\farcs7\times15\farcs7$&Uniform&$ \geq\rm0.4\,k\uplambda$&15\arcsec&55.6\\
 \hline 
 & HBA&IM19&$5\farcs5 \times 5\farcs5$&Uniform&$ \geq\rm0.4\,k\uplambda$&none&141.0\\
LOFAR$^{\dagger}$ & HBA &IM20&$8\farcs0\times8\farcs0$&Uniform&$ \geq\rm0.4\,k\uplambda$&7\arcsec&170.3\\
 (120-180 MHz)&  HBA &IM21&$15\farcs7\times15\farcs7$&Uniform&$ \geq\rm0.4\,k\uplambda$&14\arcsec&192.3\\
 \hline
\\
\end{tabular}
\begin{tablenotes}[flushleft]
  \footnotesize
   \vspace{0cm}
   \item\textbf{Notes.} Imaging was always performed using multi-scale clean, $\tt{nterms}$=2 and $\tt{wprojplanes}$=500. For all images made with {\tt Briggs} weighting we used ${\tt robust}=0$;\,$^{_\dagger}$ For data reduction steps, we refer to \cite{Rajpurohit2018} and \cite{vanWeeren2016}. 
 
    \end{tablenotes}
    \end{threeparttable} 
\label{Tabel:imaging}
\end{table*}

Recent high-resolution radio observations added new challenges as they unveiled the existence of filamentary structure within the relics on various scales \citep[e.g.][]{Owen2014,vanWeeren2017b,Gennaro2018}, including our first work on the "Toothbrush" relic \citep{Rajpurohit2018}. The origin of the  filaments is unknown. Such features are found on scales that are still prohibitive to model even with modern numerical simulations of magnetic field growth in galaxy clusters \citep[][]{review_dynamo,2019MNRAS.486..623D}. Therefore, gathering consistent radio data from them is presently crucial to understand their origin.

\section{1RXS J0603.3+4214}
1RXS\,J0603.3+4214 is a merging galaxy cluster located at a redshift of $z=0.22$. It hosts one of the largest and brightest relics, known as the Toothbrush \citep{vanWeeren2012a}. The cluster also contains two fainter relics (E and D) and a giant elongated radio halo. 

The Toothbrush relic has long been an object of intense scrutiny. 
Numerous observational studies of the Toothbrush relic have been performed across many radio frequencies \citep{vanWeeren2012a,Stroe2016,Kierdorf2016,vanWeeren2016,Rajpurohit2018}.
It shows an unusual linear morphology with a size of about 1.9\,Mpc and extends far towards the cluster center, into the low-density ICM.  The relic structure shows three distinct components resembling the brush (B1) and the handle (B2+B3) of a toothbrush. The relic morphology is enigmatic. \cite{Bruggen2012} simulated a triple merger to explain how the structures created at the shock front could extend far into the low-density ICM regime.

X-ray observations of 1RXS\,J0603.3+4214 revealed the presence of a weak shock of $\mathcal{M} \approx1.5$ at the outer edge of the brush \citep{Ogrean2013,vanWeeren2016,Itahana2017}.  In contrast, the radio observations  suggest a high Mach number shock of $\mathcal{M} \sim 3.7$  \citep{Rajpurohit2018}. There thus remains a large discrepancy between the Mach number derived from the X-ray and radio observations. 
 
The Toothbrush has also been observed at high frequencies $( >2\,\rm GHz)$, where there is still some disagreement about its integrated radio spectral shape. \cite{Stroe2016} reported interferometric observations of the Toothbrush relic from 150\,MHz to 30\,GHz. They highlighted a steepening of the integrated spectrum beyond 2\,GHz (from $\alpha\,= \,-1.00 $ to $-1.45$). In contrast, the studies conducted with high frequency single dish observations at 4.8\,GHz and 8.3\,GHz have not shown significant evidence of spectral steepening \citep{Kierdorf2016}.

We studied the Toothbrush with the VLA in L-band \citep{Rajpurohit2018}. These observations allowed us to reveal filamentary structures on various scales, likely resulting from projection effects and the magnetic field distribution in the ICM. The Toothbrush is one of the brightest relics in the sky, thus, it provides an unparalleled chance to study these detailed structures.  

Here, we present the results of further observations of the galaxy cluster 1RXS\,J0603.3+4214 with the Karl G. Jansky Very Large Array (VLA) and the upgraded Giant Metrewave Radio Telescope (uGMRT). These observations were mainly undertaken to provide higher-resolution radio emission from the Toothbrush relic, thus allowing us to study the exceptional radio emission in more detail than had been done previously. 

The layout of this paper is as follows. In Sec.\,\ref{obs}, we present an overview of the observations and data reduction. The new radio images are presented in Sec.\,\ref{radioimages}. The results obtained with the spectral analysis are described in Sec.\,\ref{relics_analysis} and Sec.\,\ref{halo}, followed by the summary in Sec.\,\ref{sec::summary}. 

Throughout this paper we assume a $\Lambda$CDM cosmology with $H_{\rm{ 0}}=70$ km s$^{-1}$\,Mpc$^{-1}$, $\Omega_{\rm{ m}}=0.3$, and $\Omega_{\Lambda}=0.7$. At the cluster's redshift, $1\arcsec$ corresponds to a physical scale of 3.64\,kpc. All output images are in the J2000 coordinate system and are corrected for primary beam attenuation.


\section{Observations and Data Reduction}
\label{obs}
\subsection{VLA observations}

The galaxy cluster 1RXS\,J0603.3+4214 was observed with the VLA in C- and S-band (project code: 17B-367 and 18B-238). VLA C-band observations were taken in C-configuration while the S-band in B and C configurations. An overview of the observations and frequency bands is given in Table\,\ref{Tabel:Tabel1}. 

Due to the large angular size of the cluster, the C-band observation was pointed on the Toothbrush relic.  The total recorded bandwidth was 4\,GHz. The 34 spectral windows, each having 64 channels, were used to cover the frequency range of 4-8\,GHz. S-band observations, pointed at the cluster center, were split into 16 spectral windows, each divided into 64 channels to cover the frequency range of 2-4\,GHz. The total recorded bandwidth for S-band observations was 2\,GHz. 

For both S- and C-band observations, all four polarization products (RR, RL, LR, and LL) were recorded. For each configuration, 3C147 and 3C138 were included as the primary calibrators, observed for 5-10 minutes each at the start of the observing run or in some case at the end of the observing run. J0555+3948 was included as a secondary calibrator and observed for $\sim5\,\rm minutes$ after every 25-35 minutes target run.

  \begin{figure*}[!thbp]
    \centering
    \includegraphics[width=1.0\textwidth]{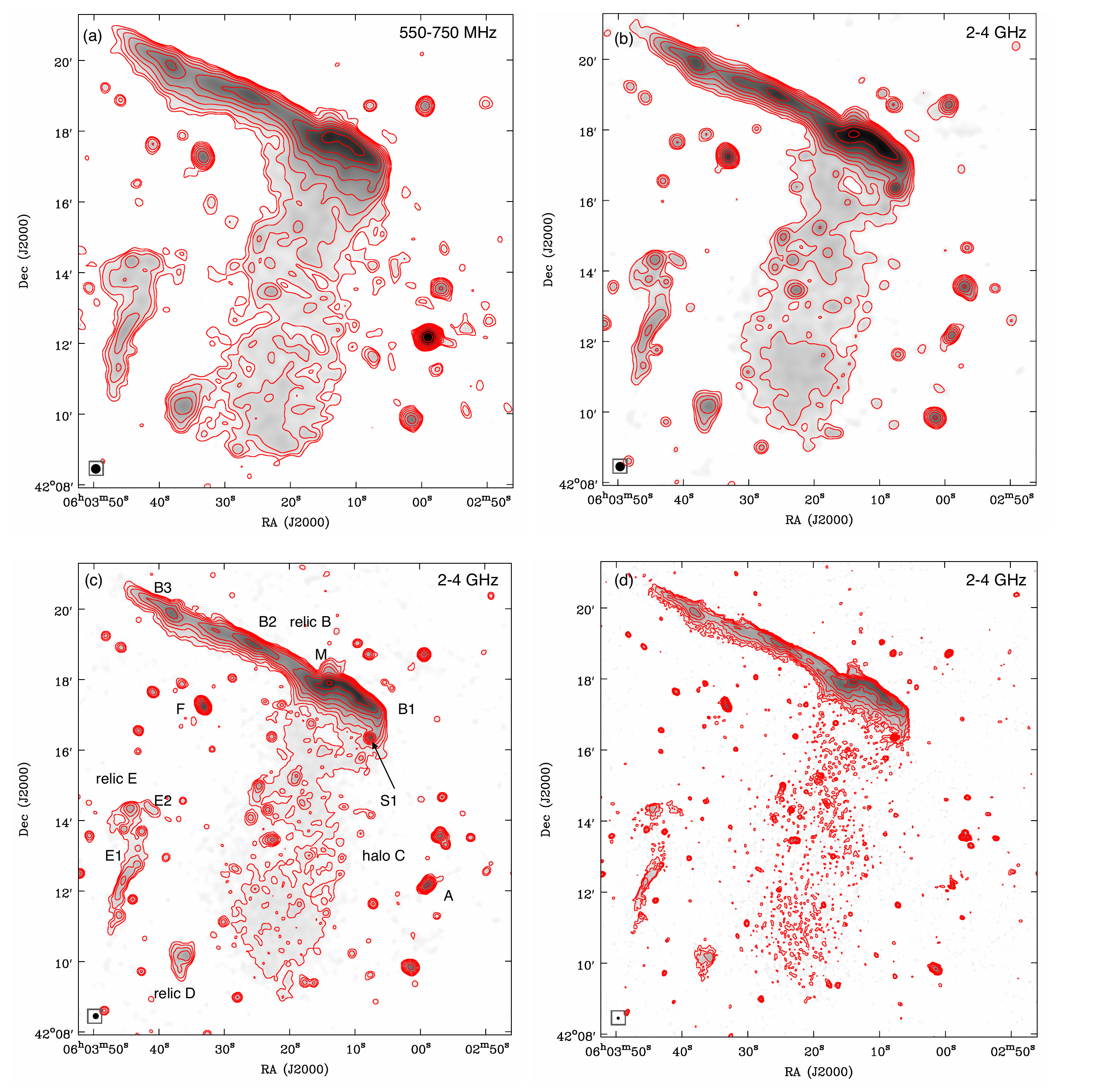}
  \vspace{-0.8cm}
    \caption{uGMRT (550-750\,MHz) and VLA S-band band images of 1RXS\,J0603.3+4214. The known diffuse emission sources, namely the bright Toothbrush, the other two fainter relics to the west and the large elongated halo are recovered in both the uGMRT and the VLA (S-band) observations. The image properties are given in Table\,\ref{Tabel:imaging}. Here, panel (a), (b), (c), and (d) correspond to IM17, IM9, IM8, and IM5, respectively. Contour levels are drawn at $[1,2,4,8,\dots]\,\times\,4\,\sigma_{{\rm{ rms}}}$. The beam size are indicated in the bottom left corner of the image.}
      \label{fig1a}
  \end{figure*}  

The data were calibrated and imaged with $\tt{CASA}$ \citep{McMullin2007} version 4.7.0. The data obtained from different observing runs were calibrated separately but in the same manner. The first step of data reduction consisted of the Hanning smoothing of the data. The data were then inspected for RFI (Radio Frequency Interference) and affected data were mitigated using ${\tt tfcrop}$ mode from the ${\tt flagdata}$ task. The low amplitude RFI was flagged using ${\tt AOFlagger}$ \citep{Offringa2010}. Next, we determined and applied elevation dependent gain tables and antenna offset positions.  We then corrected for the bandpass using the calibrator 3C147. This prevents flagging of good data due to the bandpass roll-off at the edges of the spectral windows. 

  \begin{figure*}[!thbp]
        \vspace{-0.35cm}
    \centering
    \includegraphics[width=0.95\textwidth]{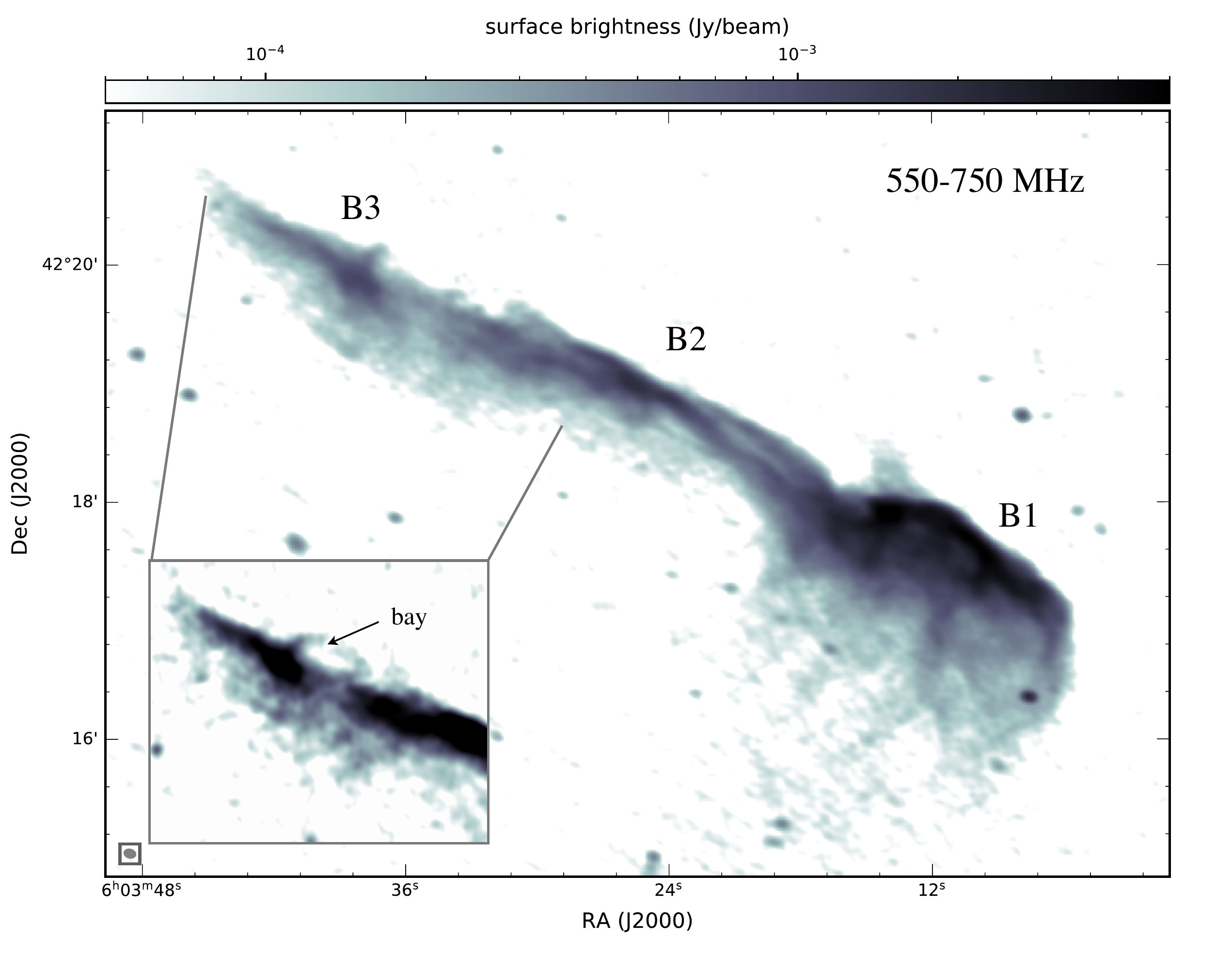}
     \vspace{-0.3cm}
    \caption{High resolution uGMRT image of the Toothbrush relic confirming the complex filamentary structures visible at 1-2\,GHz \citep{Rajpurohit2018} also in the 550-750\,MHz range. The beam size is indicated in the bottom left corner of the image.  The inset shows a low surface-brightness emission, labelled as "bay", connecting the B2 region with B3. The image properties are given in Table\,\ref{Tabel:imaging}, IM14.}
      \label{uGMRT5arcsec}
  \end{figure*}

As a next step in the calibration, we used the S- and C-band 3C147 model provided by the ${\tt CASA}$ software package and set the flux density scale according to \cite{Perley2013}. An initial phase calibration was performed using both the calibrators over a few channels per spectral window where the phase variations per channel were small. We corrected for parallel-hand delay and determined the bandpass response using the calibrator 3C147. Applying the bandpass and delay solutions, we proceeded with the gain calibration for the primary calibrators.  The calibration solutions were then applied to the target field. For all different observing runs, the resulting calibrated data were averaged by a factor of 4 in frequency per spectral window.

After calibrating each configuration, we created initial images of the target field. The imaging of the data was executed with the ${\tt CASA}$  task ${\tt CLEAN}$. For wide-field imaging, we employed the W-projection algorithm \citep{Cornwell2008} which takes into consideration the effect of non-coplanarity. To take into account the spectral behavior of the bright sources in the field, we imaged each configuration using $\tt{nterms}$=3 \citep{Rau2011}. The deconvolution was always performed using a multi-scale multi-frequency ${\tt CLEAN}$  algorithm \citep{Rau2011} and with ${\tt CLEAN}$ masks generated through ${\tt PyBDSF}$ \citep{Mohan2015}. The multi-scale setting assumes that the emission can be modeled as a collection of components at a variety of spatial scales, hence, this setting is necessary to account for the extended emission. We used the ${\tt Briggs}$ weighting scheme with a robust parameter of 0.

Following initial imaging, we performed self-calibration with a few rounds of phase-only calibration and checked that the model, to be included for self-calibration, did not have artifacts. We then ran a final round of amplitude-phase calibration.
After deconvolving each configuration independently, we subtracted the bright source A from the $uv$-data (for labels, see Fig.\ref{fig1a}.  Since the noise level increases with higher \texttt{nterms}, the subtraction of the bright source A allowed us to image the data with $\tt{nterms}$=2. To speed up imaging, we also subtracted all sources outside the cluster region in the $uv$-plane.

The final S-band deep Stokes I continuum images were made by combining the data from B and C configurations using $\tt{nterms}$=2  that resulted in a lower noise level. The CASA task ${\tt widebandpbcor}$ was used to correct for the primary beam attenuation.

\subsection{uGMRT observations }
The GMRT observation of 1RXS\,J0603.3+4214 was carried out with the upgraded GMRT (uGMRT) in band\,4 (proposal code: 34\_073). The data were recorded into 4096 channels covering a frequency range of 550-750\,MHz with a sampling time of 8 seconds. All four polarization products were recorded. The primary calibrator 3C147 was included to correct for the bandpass and phase. For polarization calibration, the secondary calibrators 3C138 and 3C286 were observed. The 5 minutes scan on each of 3C84, 3C48, and 1407+284 were also recovered during the observing session. 

The uGMRT data were calibrated with $\tt{CASA}$, version 4.7.0. We split data into 4 sets, namely set1 (channels\,$0{-}999$), set2 (channels\,$1000{-}1999$), set3 (channels\,$2000{-}2999$), and set4 (channels\,$3000{-}4096$). The data were first visually inspected for the presence of RFI, where affected data were subsequently removed using ${\tt AOFlagger}$. Initial phase calibration was performed using 3C147 and were subsequently used to compute the parallel-hand delay. The primary calibrator 3C147, together with 3C138 and 3C286, were used for flux density and bandpass calibration. We used the \cite{Perley2013} extension to the \cite{Baars1997} scale for the absolute flux density calibration. Applying the bandpass and delay solutions, we proceeded with the gain calibration for 3C147, 3C138, and 3C286. All relevant solution tables were applied to the target field.  Each set was then averaged by a factor of 4 in frequency. The averaging of the uGMRT and VLA data was done to permit performing the RM-Synthesis and QU-fitting. The polarization results of the VLA and uGMRT observations of the 1RXS\,J0603.3+4214 field will be presented in a subsequent paper (Rajpurohit et al. in prep).

\begin{figure*}
      \vspace{-0.45cm}
    \centering
    \includegraphics[width=0.496\textwidth]{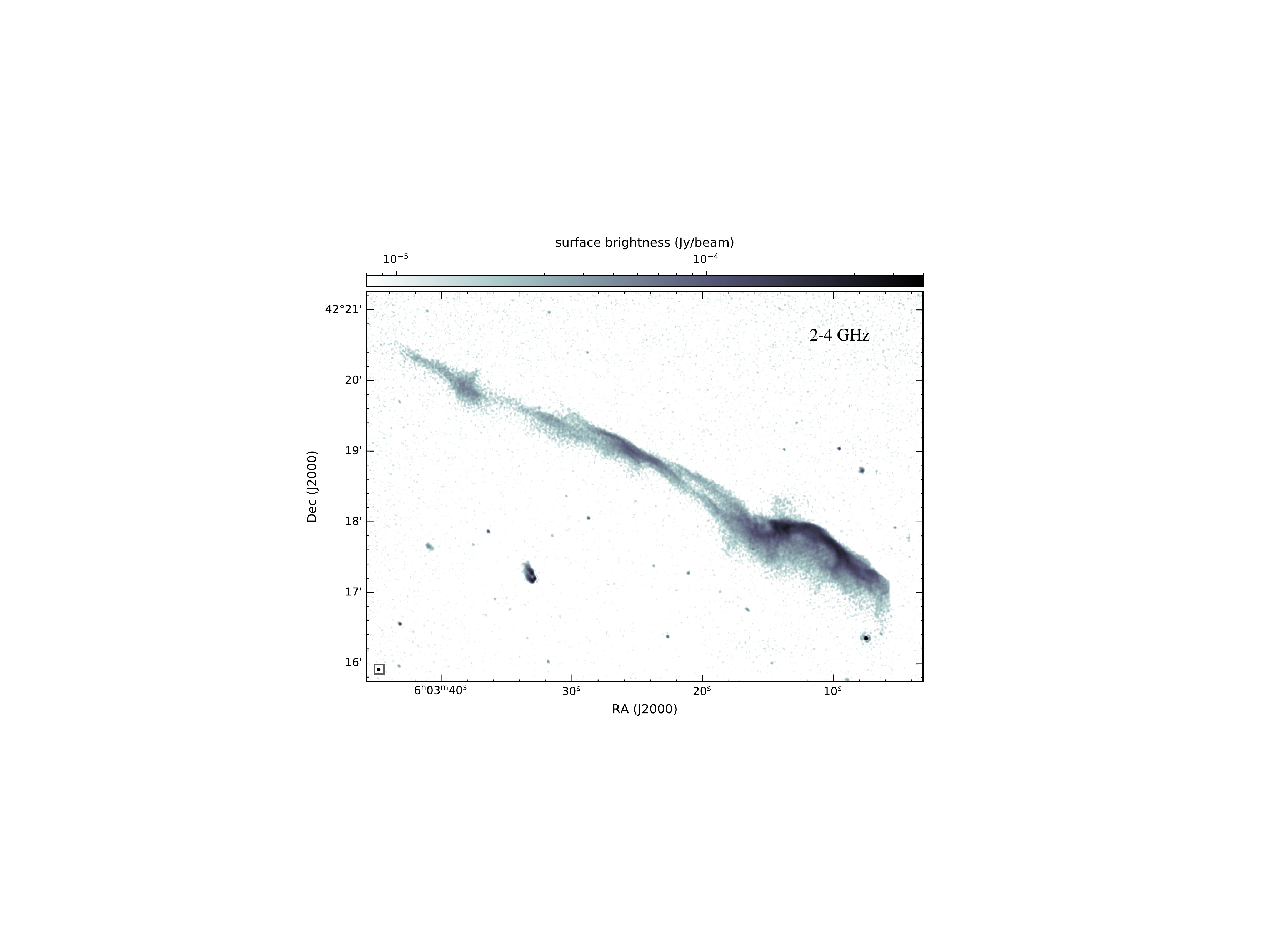}
    \includegraphics[width=0.487\textwidth]{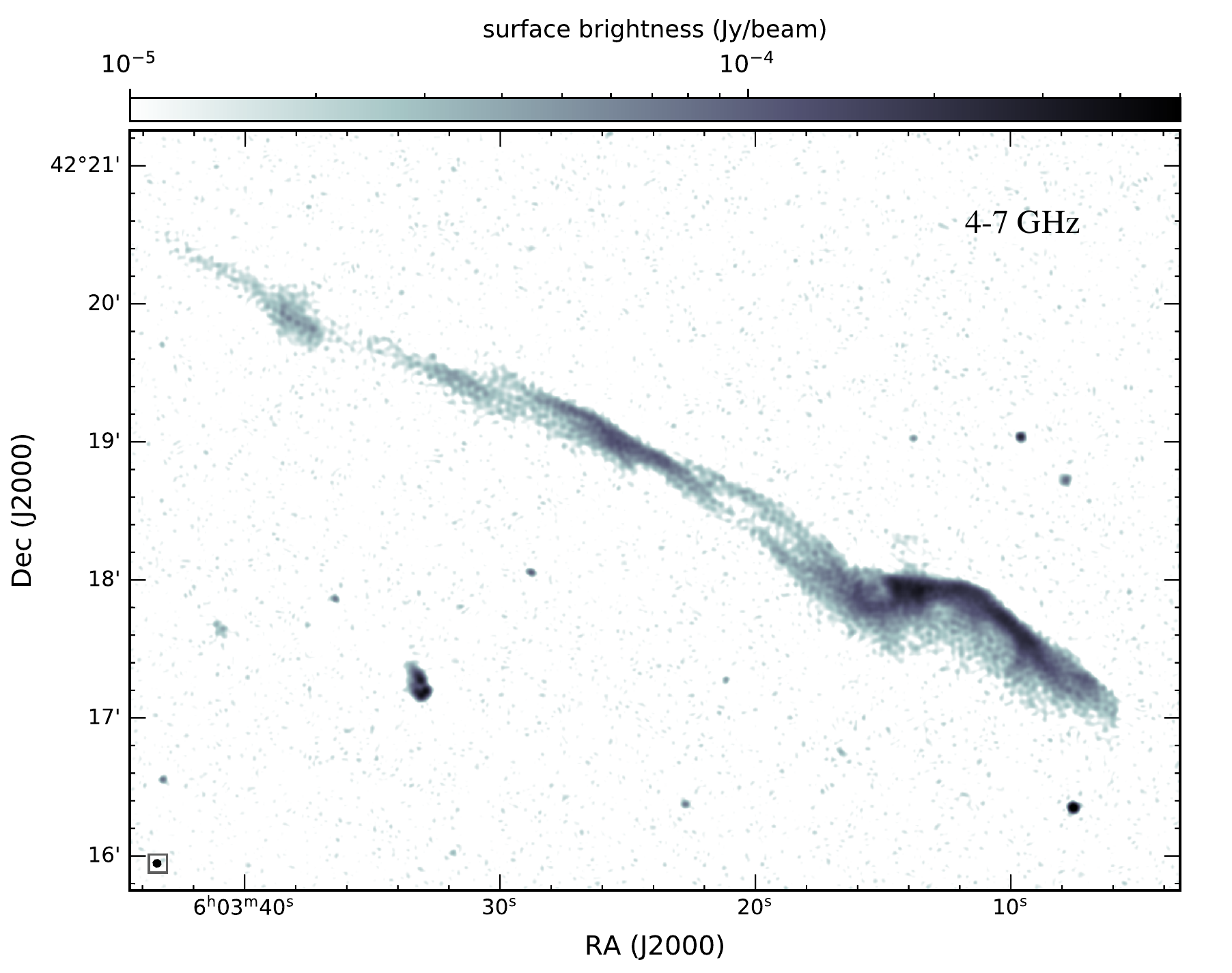}
    \caption{S-band (2-4\,GHz) and C-band (4-7\,GHz) high resolution image of the Toothbrush relic. Most of the large-scale filaments are evidently recovered up to 7\,GHz. The beam sizes are indicated in the bottom left corner of the image. The image properties are given in Table\,\ref{Tabel:imaging}, IM4 and IM1. }
      \label{fig2}
  \end{figure*}

After this, we combined all data (i.e., set1, set2, set3, and set4). Several rounds of phase self-calibration were carried out on the combined data followed by two final rounds of amplitude and phase self-calibration. We visually inspected the self-calibration solutions and manually flagged some additional data. Deconvolution was performed with $\tt{nterms}$=2, $\tt{wprojplanes}$=500, and $\tt{Briggs}$ weighting with robust parameter 0.


\section{Results: Continuum images}
\label{radioimages}

We show in Fig.\,\ref{fig1a} the resulting total intensity uGMRT (550-750\,MHz) and VLA S-band (2-4\,GHz) radio continuum images of 1RXS\,J0603.3+4214. These images are created with different uv-tapers to emphasize the radio emission on various spatial scales. The sources are labeled following \cite{vanWeeren2012a} and \cite{Rajpurohit2018}. An overview of the properties of the diffuse radio sources in the cluster is given in Table\,\ref{Tabel:Tabel4}.

\begin{figure*}
\centering
\includegraphics[width=1.0\textwidth]{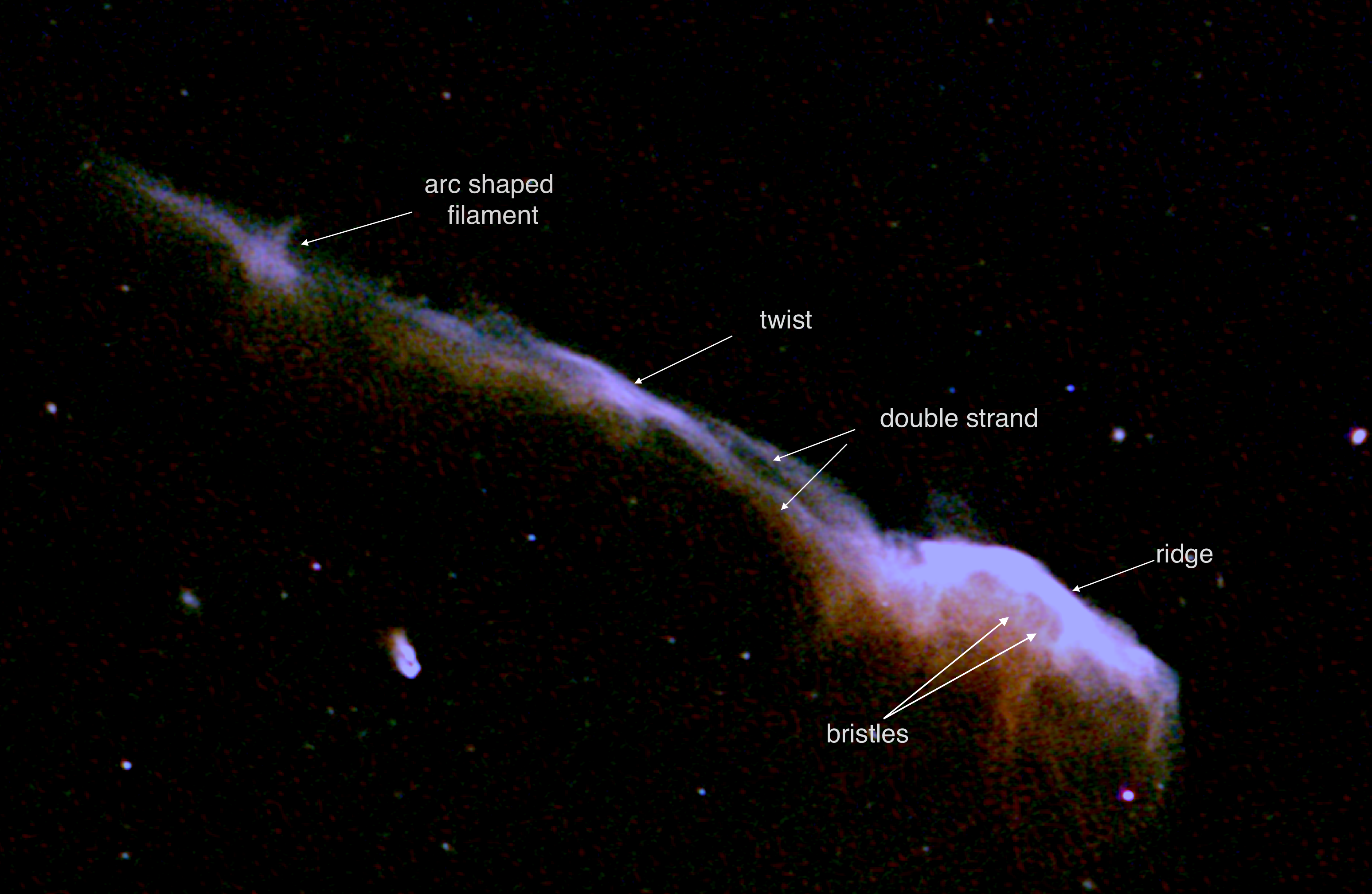}
\caption{Combined uGMRT (550-750\,MHz), VLA L-band (1-2\,GHz), and VLA S-band (2-4\,GHz) image of the Toothbrush relic manifesting variety of small and large scale complex filamentary structures. The intensity in red, green, and blue shows the radio emission observed with the uGMRT, VLA L-band and VLA S-band, respectively. The colors reflect the local spectral index variations within the relic, although there is not a 1:1 match} 
\label{fig1}
\end{figure*}

\subsection{The Toothbrush relic}
The most prominent source in the 1RXS\,J0603.3+4214 field is the large toothbrush shaped relic which is detected in all of the radio maps. The high resolution deep wideband images of the Toothbrush relic are shown from Fig.\,\ref{uGMRT5arcsec} to Fig.\,\ref{fig2}. The synthesized beams and sensitivities of these images are mentioned in Table\,\ref{Tabel:imaging}. 

Our new images manifest the variety of fine structures, as reported by \cite{Rajpurohit2018}, in the form of arcs, streams, and filaments of enhanced surface brightness. In Fig.\,\ref{uGMRT5arcsec}, we show our deep uGMRT $5\farcs0\times4\farcs8$ resolution image of the Toothbrush,  with {\tt Briggs} weighting and {\tt robust}=0. Thanks to the wideband receivers, we reached to a noise level of $7.7\,\upmu\,\rm Jy\,beam^{-1}$. The new uGMRT observations are sensitive to much lower surface brightness emission than the published 610\,MHz GMRT image \citep{vanWeeren2012a}.

The new uGMRT image reveals all small and large scale filamentary structures reported by \cite{Rajpurohit2018} also in the 550-750\,MHz range. It suggests that up to 2\,GHz the radio synchrotron emission has the same spatial distribution. In terms of morphology, our uGMRT image is comparable to the published GMRT image.

 \cite{Rajpurohit2018} reported a tentative detection of a bridge like emission feature connecting the northern edge of B2 to an arc-shaped filament in the B3 region. In the uGMRT image, the emission is evidently visible, see Fig.\,\ref{uGMRT5arcsec} lower left, however, the emission appears to be disconnected from B3. We label this emission feature as the "bay". We do not find any cluster galaxy close to the "bay".

All identified filamentary features excluding the small-scale features, namely bristles, are also visible in the VLA S- and C-band images; see Fig.\,\ref{fig2}. At all three frequencies, the northern edge of the "brush" region, B1,  is extremely sharp. We can set an upper limit to the width of the leading edge of the brush region of $2\arcsec$, which corresponds to 6.5~kpc.  The largest linear size (LLS) of the Toothbrush relic remains almost the same, i.e., $\sim 1.9\,\rm Mpc$, from 550\,MHz to 8\,GHz. However, the width decreases in the north-south direction with increasing frequency. The width of B1 at 550-750\,MHz, S-band and C-band is 530\,kpc, 327\,kpc, and 243\,kpc, respectively. This is expected for a source with a strong spectral index gradient across it.

In Fig.\,\ref{fig1}, we show a combined  uGMRT (550-750 MHz), VLA L-band (1-2\,MHz), and S-band (2-4\,GHz) image of the Toothbrush. The image shows clearer filaments. Various components of the Toothbrush are labeled as in \cite{Rajpurohit2018}. The emission in the B1 region consists of bright short filaments, labeled as "bristles",  with a width of 3-8\,kpc. We note that the widths of the "bristles" are almost the same at the 550-750\,MHz and the 1-2\,GHz. The filaments in the B2 region are generally extended over considerable distances. The brightest region in B2 is at the intersection of the double strand ("twist"). A low surface brightness emission, source M, to the northeast of B1 is visible at all three frequencies; see Fig.\,\ref{fig1a} and Fig.\,\ref{uGMRT5arcsec})

 \begin{figure}[!thbp]
    \centering
    \includegraphics[width=0.48\textwidth]{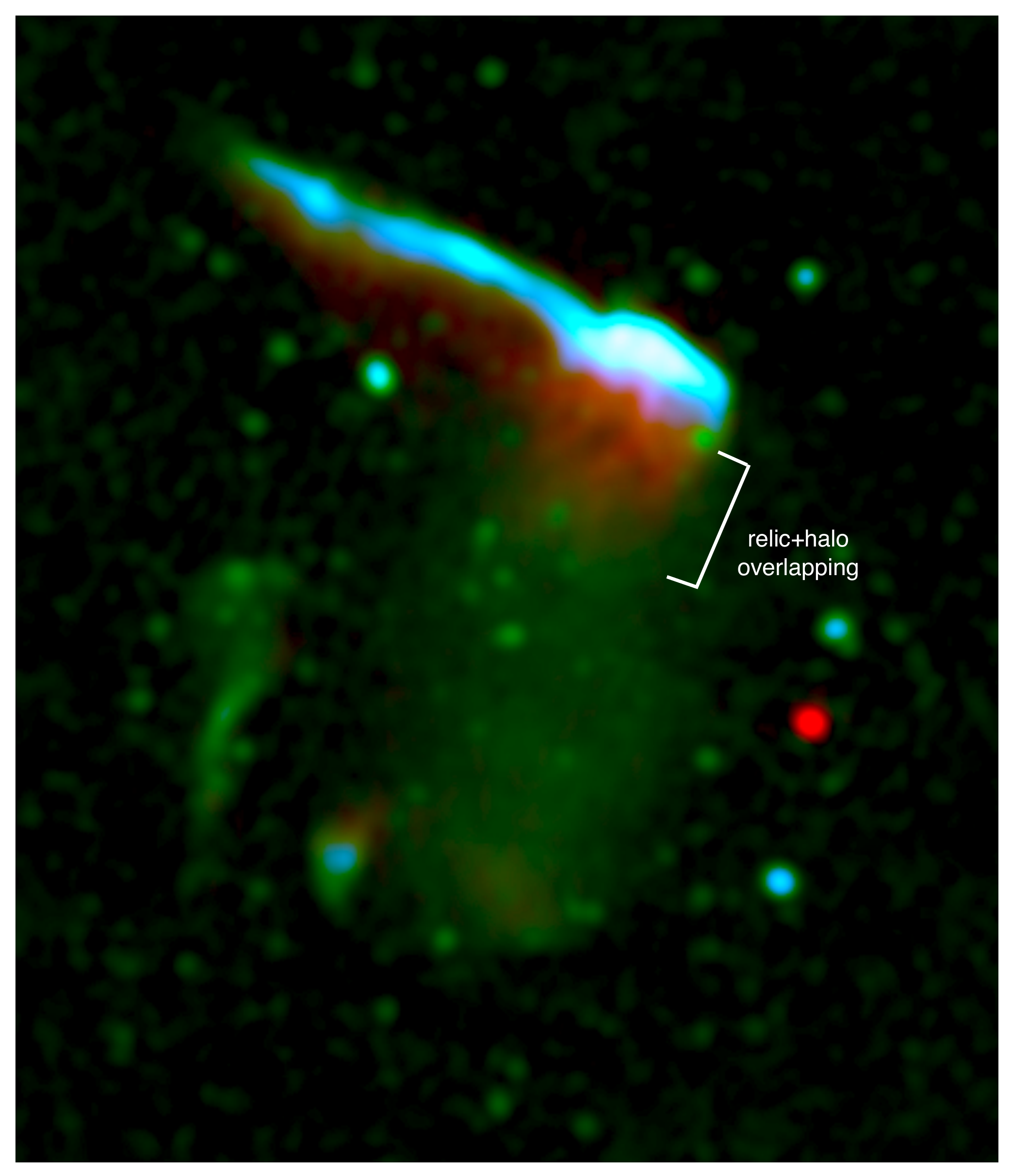}
    \caption{Combined LOFAR (120-180\,MHz), VLA L-band (1-2\,GHz), and VLA S-band (2-4\,GHz) image of the cluster at 15.7". The image  clearly shows that B1 extends out of the relic and penetrates into the halo. As discussed in Sec\,\ref{halorelic}, this is consistent with the superposition of the relic and the halo emission.  The intensity in red, green, and blue shows the radio emission observed with the LOFAR, VLA L-band, and VLA S-band, respectively. The image properties are given in Table\,\ref{Tabel:imaging}, IM10, IM13, and IM21.}
 \label{halorelicoverlap}
 \end{figure}

Our high-resolution radio maps show filamentary features on various scales that are visible from 550\,MHz to 8\,GHz. In fact, radio relics when imaged at high resolution mostly show filamentary substructure. The origin of filaments that we discovered is challenging to explain. The observed filaments could be projections of substructures from a complex shaped shock front or may reflect variations of the electron acceleration efficiency and trace a fluctuating Mach number distribution internal to relics \citep{wi17}. Also, they may illuminate regions where different magnetic field domains are forced to merge \citep{Owen2014}. 

\subsection{Relics E and D}
The other two fainter relics E and D are well detected in the S-band images, Fig.\,\ref{fig1a}. The LLS of both relics are similar to those observed below 2\,GHz.

The relic E consists of two parts, E1 and E2. They are morphologically different.   The E2 region of the relic E is fainter and more extended than E1. There are recent observational examples which seem to show a connection between relics and active galactic nuclei (AGN), i.e, they provide evidence that the shock fronts re-accelerate CRe of a fossil population \citep{Bonafede2012a,vanWeeren2017a,vanWeeren2017b,Gennaro2018,Stuardi2019}. We note that several radio sources are embedded in the relic E, however, we do not find  any obvious connection to the relic.  
\setlength{\tabcolsep}{14pt}
\begin{table*}[!htbp]
\caption{ Properties of the diffuse radio emission in the cluster 1RXS J0603.3+4214 .}
\centering
\begin{threeparttable} 
\begin{tabular}{ c | c | c | c | c | c | c | c  }
\hline \hline
\multirow{1}{*}{Source} & \multicolumn{3}{c|}{VLA} & \multirow{1}{*}{uGMRT} & \multirow{1}{*}{LOFAR} &\multirow{1}{*}{$\rm LLS^{\dagger}$} & \multirow{1}{*}{$\alpha$} \\
 \cline{2-4} 
&C-band & S-band & L-band & &   &  &\\
  \cline{2-4} \cline{5-6}
  &$S_{{\rm{6\,GHz}}}$ & $S_{{\rm{3.0\,GHz}}}$ & $S_{{\rm{1.5\,GHz}}}$&  $S_{{\rm{650\,MHz}}}$ & $S_{{\rm{150\,MHz}}}$  & \\
 &mJy & mJy & mJy & mJy & mJy  &Mpc & \\
  \cline{2-3} \cline{3-4}\cline{5-6}
  \hline  
relic\,B &$68\pm2$& $138\pm4$ & $310\pm21$&$752\pm78$& $4428\pm423$&$\sim1.9$ &$-1.16\pm0.02$ \\ 
halo\,C &- & $10\pm1$ &$33\pm3 $&$48\pm5$& $490\pm56$& $\sim1.6^{\ddagger}$ &$-1.16\pm0.04$ \\ 
relic\,D &- & $1.4\pm0.1$ &$5\pm1$  &$15\pm2$ & $96\pm12$&  $\sim0.3$ & \\
relic\,E & -&$3.1\pm0.3$& $12\pm1$& $26\pm3$ &$135\pm19$& $\sim1$ & \\
region\,S &- &$1.9\pm0.1$ &$9\pm1$&$23\pm3$& $141\pm15$ & $\sim0.7$&\\
\hline 
\end{tabular}
\begin{tablenotes}[flushleft]
\footnotesize
\item{\textbf{Notes.}} Flux densities were measured from 25\arcsec resolution images, created with uniform weighting and without any uv-cut. The regions where the flux densities were extracted are indicated in the left panel of Fig.\,\ref{region}. We assume an absolute flux scale uncertainty of 10\% for the GMRT and LOFAR data, 4\% for the VLA L-band, and 2.5\% for the VLA S-, C-band data.  The flux density of the halo excludes the ``relic+halo" and region\,S; The spectral index values are obtained from images with an uvcut; see Sec.\,\ref{relics_analysis}; $^{\dagger}$the largest linear size at 3\,GHz; $^{_\ddagger}$ size of the halo includes the "relic+halo" and region\,S.
\end{tablenotes}
\end{threeparttable} 
\label{Tabel:Tabel4}   
\end{table*} 

\subsection{Radio halo}

Our new observations provide the first detailed image of the central halo emission in 1RXS\,J0603.3+4214 at 2-4\,GHz. The halo emission is best highlighted in the low resolution images, see Fig.\,\ref{fig1a}. The total extent of the halo emission is about 8\arcmin, corresponding to 1.7\,Mpc at 3\,GHz. It has low surface brightness ($> 0.03\rm \,mJy\,beam^{-1}$) and is extended along the merger axis. 

The halo is also recovered in the 550-750\,MHz uGMRT images, see Fig.\,\ref{fig1a} panel (a). The overall morphology and extent of the halo at 550-750\,MHz and 2-4\,GHz are similar to that observed by \cite{vanWeeren2016} and \cite{Rajpurohit2018} at 150\,MHz and 1.5\,GHz, respectively. However, we note that at each frequency, the transition from the relic to the halo occurs at somewhat different locations. Towards high frequencies (>1.5\,GHz), we find that there is a slight decrease in the surface brightness, apparently separating the relic and the halo emission; see Fig.\ref{fig1a} panel (d). 

In Fig.\,\ref{halorelicoverlap} we show a combined LOFAR (120-180 MHz), VLA L-band (1-2\,GHz), and VLA S-band (2-4\,GHz) image. The image nicely show that the brush region of the Toothbrush extends and penetrates into the halo. We labelled this region as "relic+halo". We will argue in Sec\,\ref{halorelic} that there is an overlap between the relic and the halo emission. 

\textit{Chandra} observations of 1RXS\,J0603.3+4214 revealed a weak shock of $\mathcal{M}\sim1.8$ at the southern edge of the halo \citep{vanWeeren2016}. We denoted this region as "region S" in Fig.\,\ref{region}. It has been reported that the region S is possibly connected to the southern shock front, thus may have a different origin \cite{Rajpurohit2018}.

Higher resolution L-band images by \cite{Rajpurohit2018} showed that there are around 32 discrete radio sources in the area covered by the halo emission. Most of these points sources are also visible at 2-4\,GHz.
The flux density of the halo, including discrete sources, at 610\,MHz and 3\,GHz are $S_{610\,\rm MHz}=141\pm17\,\rm mJy$ and $S_{\rm 3\,GHz}=18\pm2\,\rm mJy$, respectively. This is the flux density for the entire halo, including the relic+halo and region S. The measured flux density at 610\,MHz flux is significantly higher than the published narrow band GMRT observations.

\begin{figure*}[!thbp]
\vspace{-0.5cm}
    \centering
    \includegraphics[width=1\textwidth]{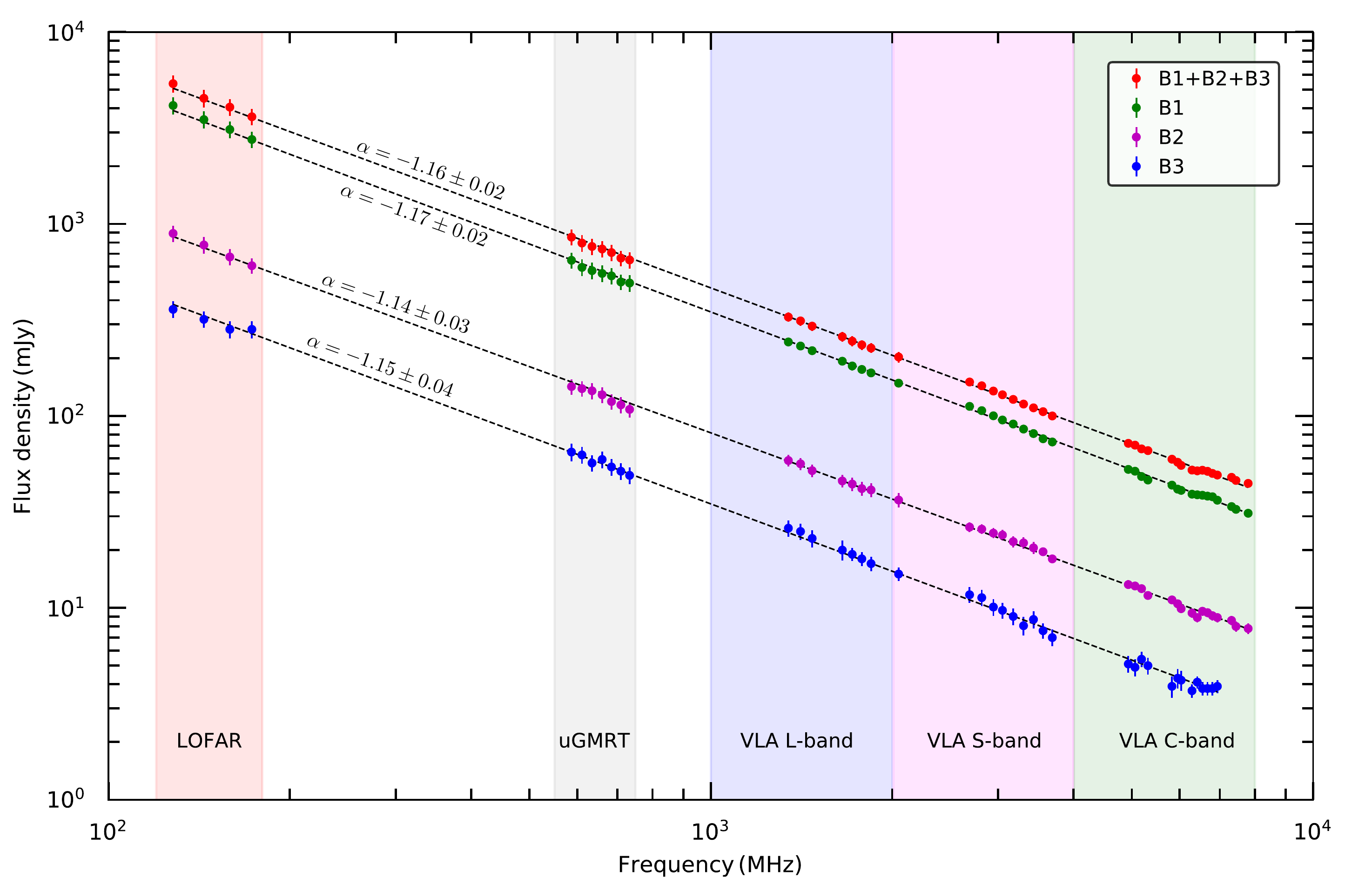}
        \vspace{-0.5cm}
    \caption{\textit{Left}: Integrated spectra of the main Toothbrush and subregions between 150\,MHz and 8\,GHz. Dashed lines are fitted power-law. The spectrum of the main Toothbrush is well described by a single power-law with slope $\alpha=-1.16\pm0.02$. We rule out any possibility of spectral steepening of the relic emission at any frequency below 8\,GHz. Despite being located at different distances from the ICM, the B1, B2, and B2 spectra are remarkably identical. The flux densities are measured from $8\arcsec$ resolution images, created using uniform weighting with a uv-cut at $0.4\rm\,k\,\uplambda$. The regions where the flux densities were extracted are indicated in the left panel of Fig.\,\ref{region}.} 
  \label{spectrum_cc}
\end{figure*} 


\section{Analysis of relics}
\label{relics_analysis}

To study the spectral characteristics of relics B, D, and E over a wide range of frequencies, we combined the deep VLA (2-8\,GHz) and uGMRT (550-750\,MHz) observations presented here with the previously presented ones at 1-2\,GHz \citep{Rajpurohit2018} and 150\,MHz \citep{vanWeeren2016}. In fact, the Toothbrush is the first relic where the radio interferometric observations over such a wide frequency range are available.  

\subsection{Flux density measurements}
\label{relic_flux}

In order to derive reliable flux densities and spectral index maps, we imaged each data set with uniform weighting and convolved the final images to the same resolution. However, such images can only be produced if we have the same uv-coverage in each interferometric observation otherwise this results in a bias in the total flux density measurements of very extended sources.

The radio observations reported here are performed with different interferometers, each of which have different uv-coverages. The shortest   baseline for the LOFAR, uGMRT, VLA L-band, VLA S-band, and VLA C-band data are $0.03\,\rm k\uplambda$, $0.2\,\rm k\uplambda$, $0.2\,\rm k\uplambda$, $0.3\,\rm k\uplambda$, and $0.4\,\rm k\uplambda$, respectively. To have the same spatial scale at all frequencies, we create images with a common lower uv-cut at $0.4\rm\,k\uplambda$. Here, $0.4\rm\,k\uplambda$ is the well sampled baseline of the uGMRT and the VLA C-band data. This uv-cut is applied to the LOFAR and the VLA L- and S-band data. To reveal the spectral properties of different spatial scales, we tapered the images accordingly. 

To measure flux densities, we create images at $8\arcsec\times8\arcsec$ resolution with a common inner uv-range mentioned above. We chose 8\arcsec resolution images for measuring the flux densities to: (1) avoid contamination from the halo emission in the "relic+halo" region. As mentioned by \cite{Rajpurohit2018} between 150-650\,MHz, the "relic+halo" is dominated by B1 emission but in contrast between 1-4\,GHz by the halo emission. Therefore, when determining the integrated spectrum this region is considered as the part of the toothbrush relic. The reason is, at $8\arcsec$ resolution the low surface brightness halo emission does not contribute much to the relic B emission and will not significantly affect our flux density measurements; (2) since it increases the signal-to-noise radio in low surface brightness regions, in particular for B3 at high frequency, namely in C-band.

Radio relics usually contain a number of discrete sources, thus making an accurate measurement of the integrated spectra difficult. The contamination by discrete sources, therefore, needs to be subtracted from the total diffuse emission first. Luckily, the Toothbrush is not in close-by to any unrelated sources except source S1, see Fig.\,\ref{fig1a} panel (c). We do not subtract its flux contribution because the source is rather faint ($S_{\rm 1.5\,GHz}= 1.8\pm0.1\rm m\,Jy$), therefore, the contamination is not relevant. 

The uncertainty in the flux density measurements are estimated as: 
\begin{equation}
\Delta S =  \sqrt {(F \cdot S_\nu)^{2}+{N}_{{\rm{ beams}}}\ (\sigma_{{\rm{rms}}})^{2}}
\end{equation}
where $F$ is an absolute flux calibration uncertainty, $S_\nu$ is the flux density, $\sigma_{{\rm{ rms}}}$ is the RMS noise and $N_{{\rm{beams}}}$ is the number of beams. We assume absolute flux uncertainties of 4\,\% and 2.5\,\% for the VLA L-band and S-, C-band \citep{Perley2013}, respectively. 
While for the uGMRT and the LOFAR data, assumed flux uncertainties is 10\,\% \citep{Chandra2004,vanWeeren2016}.

We compare our flux density measurements with values obtained with single dish observations that do not resolve out the flux density of extended sources. At 4.8\,GHz, we measure a flux density of $83\pm8\rm\,mJy$ which is higher than the value measured with Effelsberg at 4.85\,GHz, namely $68\pm 5\rm\,mJy$ \citep{Kierdorf2016}. 
However, they noted that their value is possibly too low because of the insufficient size and low quality of the radio map.  At 8\,GHz, we measure a flux density of $52\pm5\,\rm mJy$,  consistent with the value obtained with the Effelsberg telescope at 8.35\,GHz, namely $58\pm7\,\rm mJy$ \citep{Kierdorf2016}. Our interferometric high frequency measurements are in agreement with the results from the single dish telescope. This clearly indicates that our interferometric data is not affected by missing short baselines.


\subsection{Integrated radio spectra of the Toothbrush}
\label{int_relic}

Detailed studies of the integrated spectra of radio relics, over a broad range of frequencies, serve as a useful measure of the energy distribution of the relativistic electrons. The spectra provide insightful information to discriminate between competing models of particle acceleration currently proposed for radio relics.

\begin{figure*}[!thbp]
    \centering
     \includegraphics[width=0.49\textwidth]{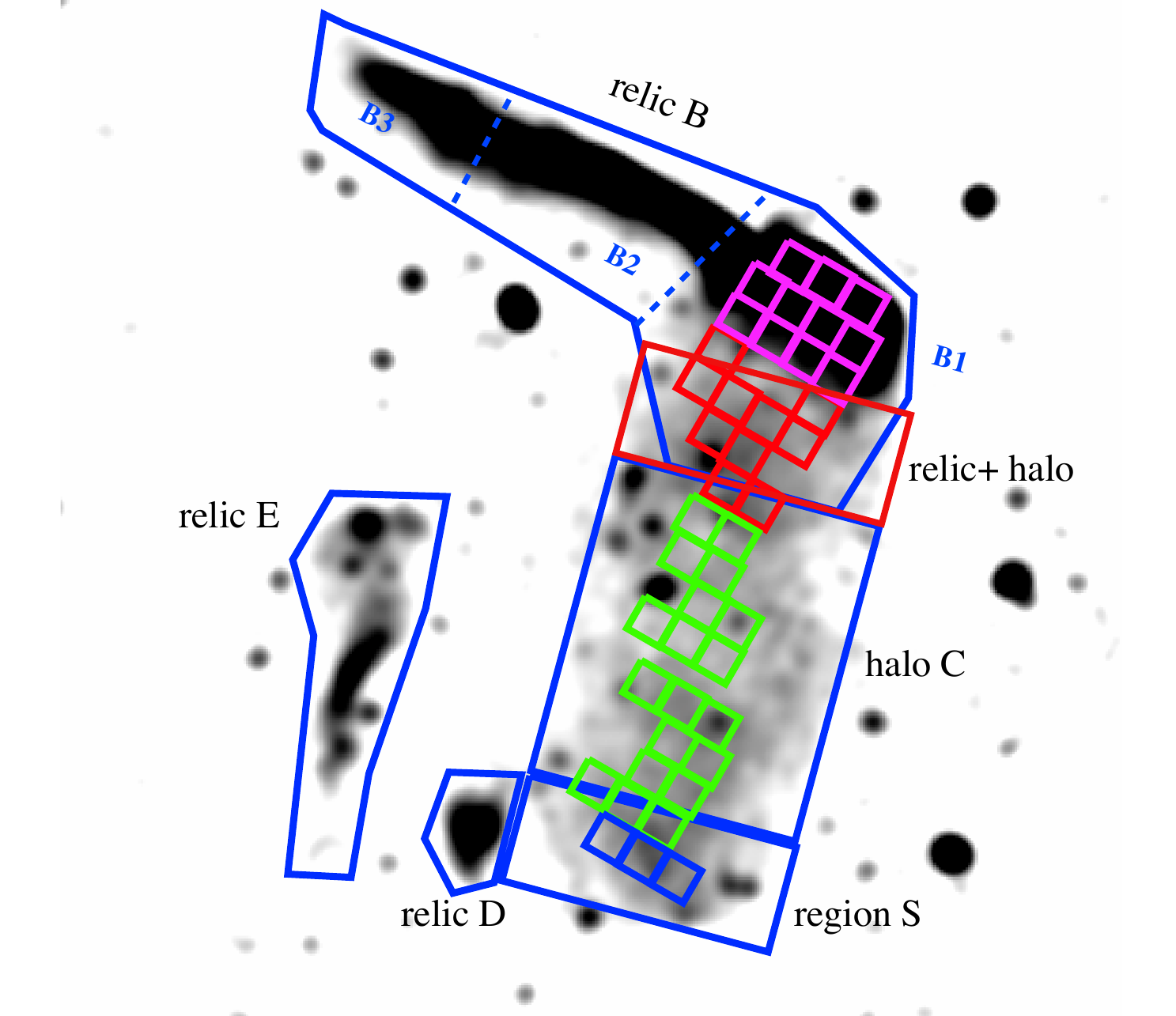}
     \includegraphics[width=0.50\textwidth]{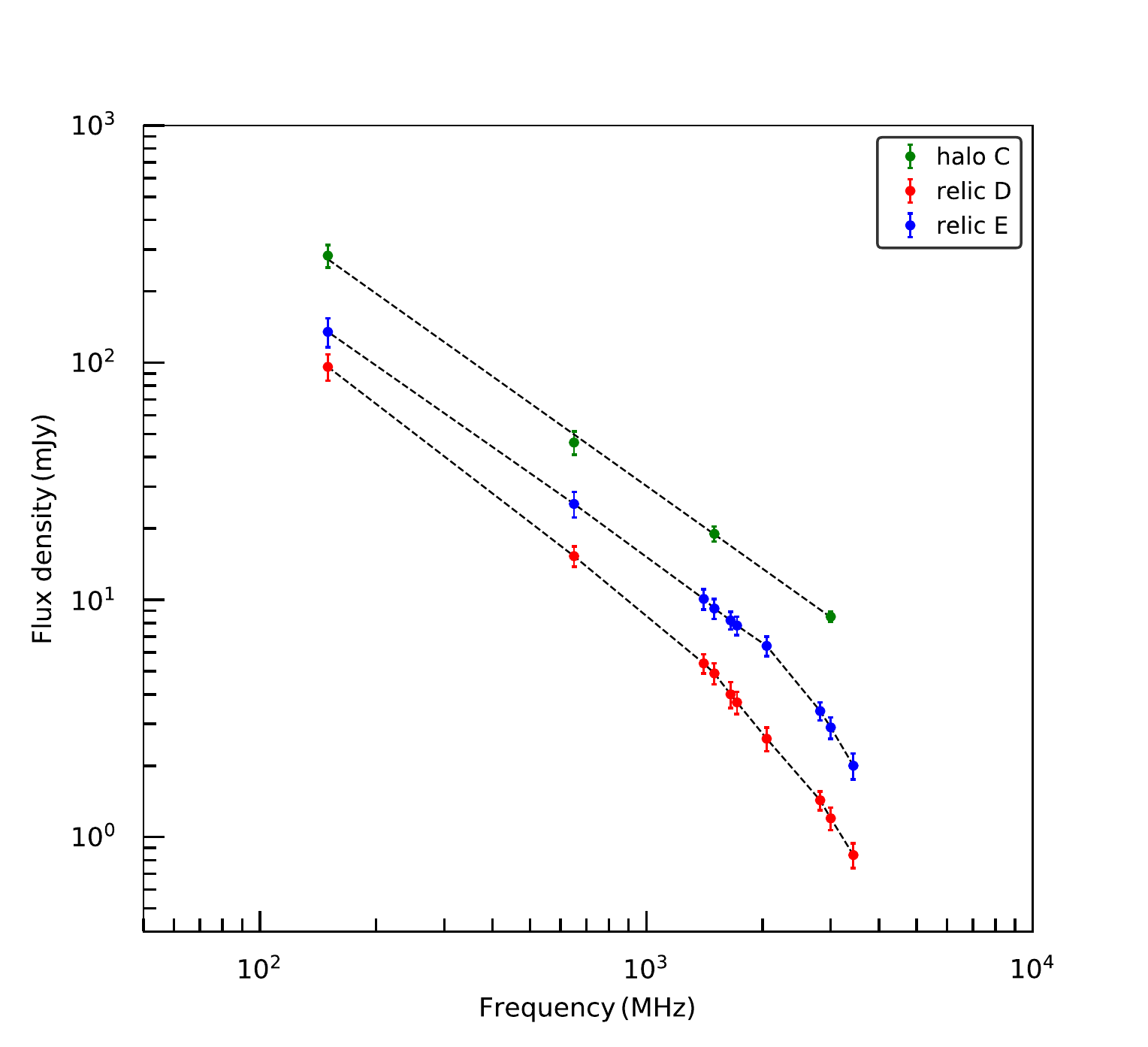}    
      \vspace{-0.2cm}
      \caption{\textit{Left}: VLA 15\arcsec resolution image depicting the regions where the integrated flux densities were measured. \textit{Right}: Integrated spectra of the halo C and relics E and  D. The spectrum of the halo C is well described by a single power-law spectrum. Relics E and D show spectral steepening beyond 1.5\,GHz. The flux densities are measured from $15\arcsec$ resolution images, created using uniform weighting with a uv-cut at $0.4\rm\,k\,\uplambda$. }
  \label{region}
\end{figure*}

The broadband interferometric observations in several frequency bands allow us to conduct the most sensitive study of the integrated spectrum of any relic up to date. To obtain the integrated spectra, we measure the flux density of the entire Toothbrush as well as in three subareas B1, B2, and B3. The regions where the flux densities were extracted are indicated in the left panel of Fig.\,\ref{region}. We use radio maps described in Sec.\,\ref{relic_flux} for all flux densities measurement of the relics. 
 
The most important and striking result of our analysis is shown in Fig.\,\ref{spectrum_cc}. We find that the integrated spectrum of the Toothbrush indeed follows a close power law over close to two decades in frequency. Our results show for the first time that not only the main Toothbrush but also the subregions B1, B2, and B3  exhibit nearly perfect power laws.

The power-law spectrum of the Toothbrush up to 8\,GHz differs from the result of an earlier study by \citet{Stroe2016}. They reported radio interferometric observations of the Toothbrush from 150\,MHz to 30\,GHz and highlighted the steepening of the integrated spectrum beyond 2\,GHz (from $\alpha= -1.00 $\,to\,$-1.45$). We note that the steepening in the flux density spectrum around 2.5\,GHz was claimed on the basis of 16\,GHz and 30\,GHz radio interferometric observations.  We find no evidence of a deviation from the power law and thus rule out the possibility of steepening found by \citet{Stroe2016}, at least below 8\,GHz.

The integrated relic emission between 150\,MHz to 8\,GHz, has a spectral index of $-1.16\pm0.02$. The integrated spectral index of the Toothbrush is consistent with our previous value of $-1.16\pm0.02$ \citep{Rajpurohit2018}. Even more astounding, we find that the integrated spectral indices of B1, B2 and B3 are remarkably close to each other, namely $\alpha_{\rm B1}=-1.17\pm0.02$, $\alpha_{\rm B2}=-1.14\pm0.03$, and $\alpha_{\rm B3}=-1.15\pm0.04$.  It is surprising that B1, B2 and B3, despite being located in regions with very different ICM densities, show spectra with an almost identical slope.

According to DSA theory in the test-particle regime, a shock of Mach number $\cal M$ generates a population of relativistic electrons with a power-law distribution in momentum \citep[see e.g.][]{Blandford1987}. In a quasi-stationary situation, including radiative losses leads to an overall radio spectrum which is again a power law. The spectral index $\alpha_{\rm int}$ of the overall --or `integrated'-- spectrum is related to the Mach number according to
\begin{equation}
  \mathcal{M} 
  = 
  \sqrt{
       \frac{\alpha_{\rm int}-1}
            {\alpha_{\rm int}+1}  } . 
 \end{equation}
If the spectral index of the regions B1, B2, and B3 reflects the Mach number of the shock, our observations indicate that the Mach number of the shock is remarkably uniform along the Toothbrush. We present a qualitative comparison with a recent numerical simulation of radio relics in Sec.\,\ref{simulation_part}. 

Other than the Toothbrush, integrated radio spectra over a wide range of frequencies are available only for five radio relics; the Sausage relic, the relic in A2256, Bullet cluster, ZwCl 0008.8+5215, and A1612 \citep{trasatti2015,Stroe2016,Kierdorf2016}. Out of these five mentioned relics, a spectral steepening has been detected between 2-5 GHz for three relics, namely the Sausage relic, the relic in A2256, and Bullet cluster. In contrast to this, some authors reported no steepening in the integrated spectrum for the Sausage relic \citep{Kierdorf2016,Loi2017}. 

There are several reasons that may have led to the these differing results. The integrated spectra of these relics are derived from combining single-dish observations at high frequency (mostly above about 4\,GHz) and interferometric observations at low frequencies (mostly below about 1.4 GHz). While interferometric measurements might underestimate the flux density as a result of missing short spacings, single-dish measurements lack resolution that might result in overestimate of the flux density owing to source confusion. Hence, it remains unclear if the integrated spectrum of these three relics are curved at high frequencies or not.  

To derive the integrated spectrum of the relic E, we first subtract the flux density contribution from several discrete sources embedded in the relic. The resultant spectrum is shown in the right panel of Fig.\,\ref{region}, along with that of relic D. Unlike the Toothbrush, the overall spectra of relics E and D  steepen at high frequencies.


\subsection{Comparison with numerical simulations}
\label{simulation_part}

At face value, it seems surprising that a single Mach number can characterize the entire shock surface across such a large distance. Numerical studies have indeed consistently reported that at least at the resolution probed by recent simulations and within their rather simplistic physical model, a rather broad distribution of shock strength is expected for merger shocks \citep[e.g.][]{Hoeft2011, Skillman2013, 2018ApJ...857...26H, 2019arXiv190911329W}.

\begin{figure}[!thbp]
    \centering
    \includegraphics[width=0.49\textwidth]{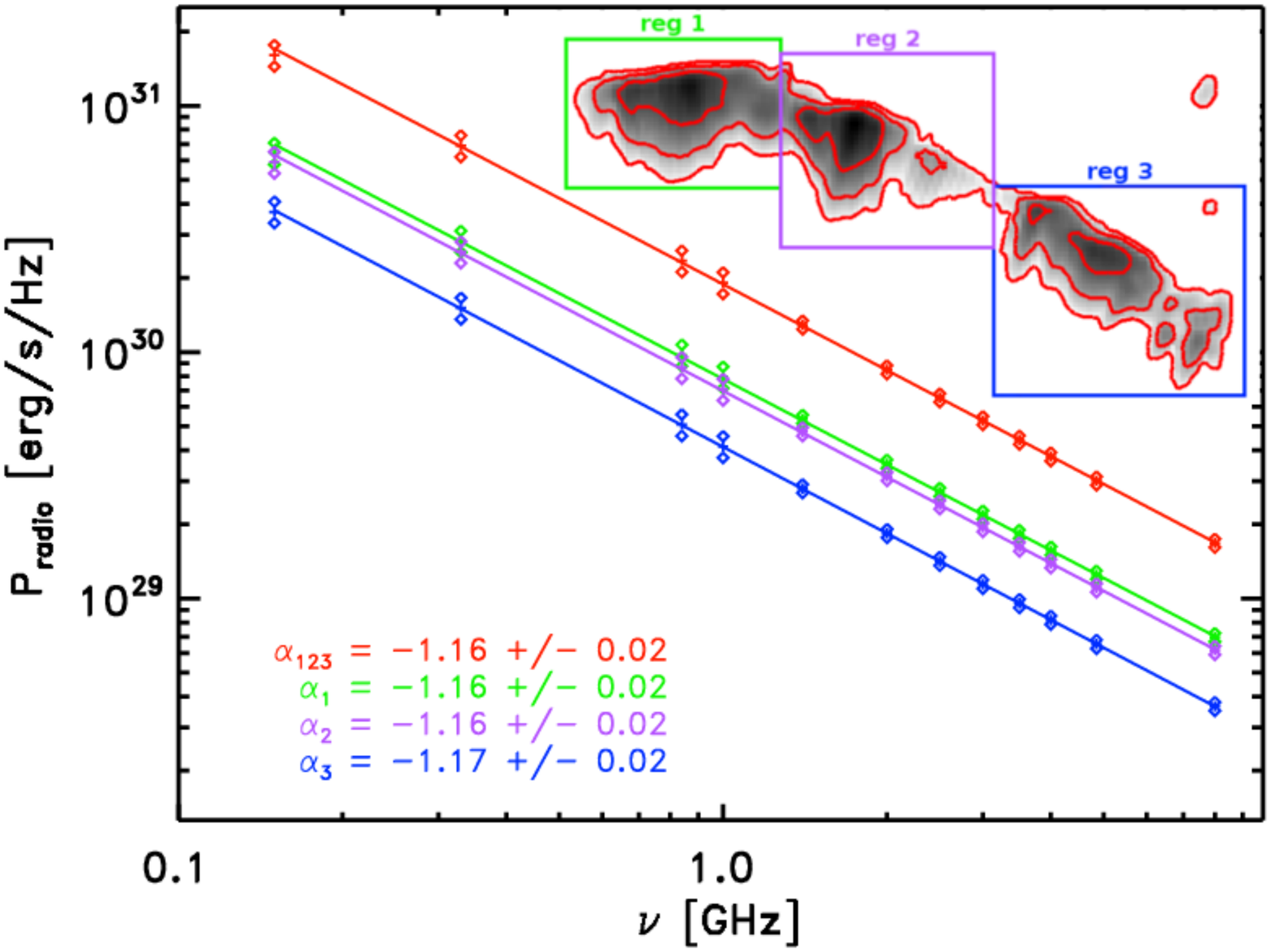}
  \vspace{-0.5cm}
    \caption{Integrated spectra for the radio relic in a recent cosmological numerical simulation \citep[][]{2019arXiv190911329W}. The red diamonds give the integrated spectrum of the whole relic, while the other diamonds give the spectra for the three subregions as in the inset (the color map gives the total intensity of the radio emission at $1.4$\,GHz. The simulated relic is $\approx 1.2$ Mpc in size.}
  \label{relic_denis}
\end{figure}

To investigate this issue in more detail, we computed the integrated emission spectra for a simulated radio relic, recently studied by \citet{2019arXiv190911329W}. We remark that this simulation is not meant to reproduce the real Toothbrush relic, as the morphology of the simulated emission, see Fig. \ref{relic_denis}, and the properties of the host cluster are different from the 1RXSJ0603.3+4214. However, \citet{2019arXiv190911329W} presented the most detailed modeling of radio relic emission to date in numerical simulations, using an ENZO-MHD simulation with a maximum resolution of $\approx 4$\,kpc \citep[][]{2019MNRAS.486..623D}. Unlike in previous studies, it models not only the injection of electrons by DSA, but also the time-dependent downstream cooling behind the shock edge.

 \begin{figure*}
    \centering
    \includegraphics[width=0.49\textwidth]{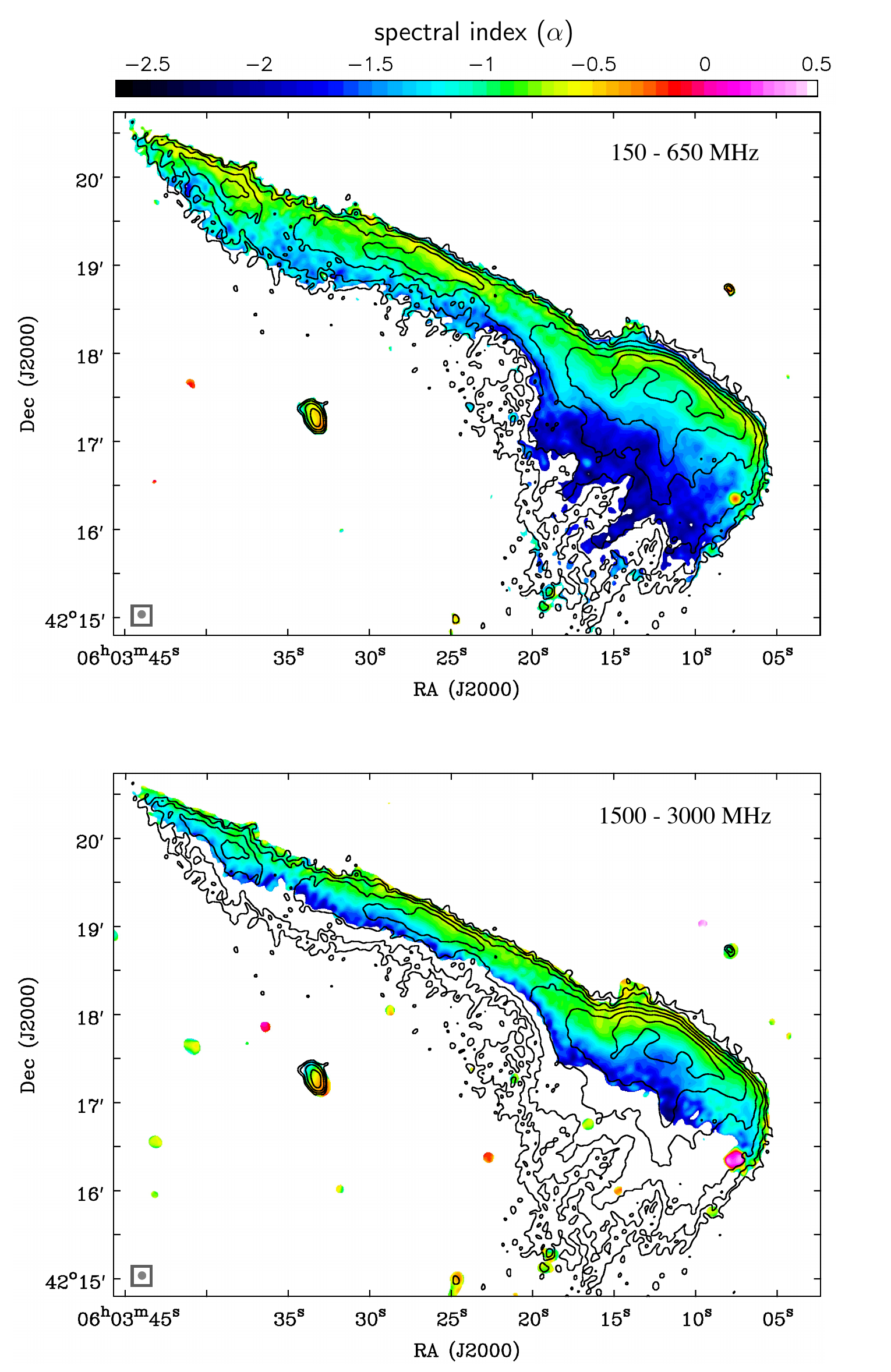}
    \includegraphics[width=0.49\textwidth]{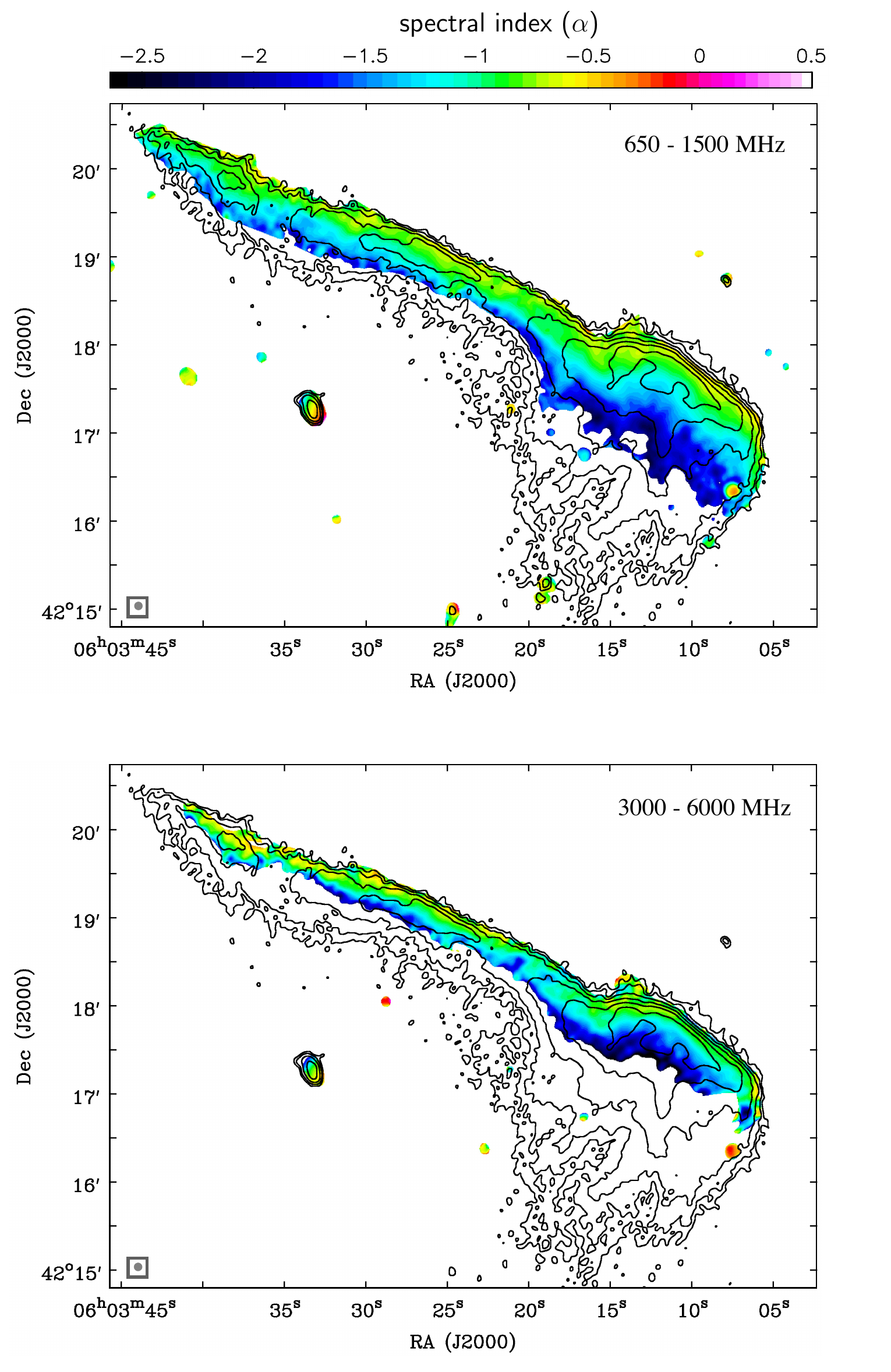}
          \vspace{-0.3cm}
    \caption{Spectral index maps of the Toothbrush, at $5\farcs5$ resolution, showing remarkably similar spectral index patterns up to 6\,GHz. The variations in the spectral index at the outer edge
along the relic is clearly visible, reflecting inhomogeneities in the injection index. \textit{Top:} low-frequency spectral index maps between 150 to 1500\,MHz. \textit{Bottom:} high-frequency spectral index maps between 1.5 to 6\,GHz.  The color bar shows spectral index $\alpha$ from -2.5 to 0.5. Bottom: In all maps, contour levels are drawn at $[1,2,4,8,\dots]\,\times\,4\,\sigma_{{\rm{ rms}}}$ and are from the 150\,MHz LOFAR image. These maps were created using the IM2, IM6, IM11, IM15 and IM19 (see Table\,\ref{Tabel:imaging} for image properties).}
      \label{toothbrush_spectral_index}
  \end{figure*}

The simulated relic forms in a $M_{\rm 100}\sim 10^{15} M_{\odot}$ galaxy cluster after a major merger at $z \approx 0.2$. It is located at a distance of $\sim 1.7\,\rm Mpc$ from the cluster center and is $\approx 1.2$ Mpc long (extending $\approx 500$ kpc along the line of sight). The 3-dimensional distribution of the Mach numbers at the shock is broad, namely $2 \leq \mathcal{M}\leq 4$. The magnetic field strength at the relic is in the range $0.1~ \rm \upmu G\leq B \leq 2~ \rm \upmu G$, with a radio-weighted mean value of $\approx 1.05~ \rm \upmu G$. The total radio power of the relic, including the effect of shock re-accelerated electrons, is $\approx 10^{30}\,\rm erg\,s^{-1}\,Hz^{-1}$ at $1.4$\,GHz. Thus, we can sum over all the simulated emission at different frequencies, to derive a simulated spectral distribution expected from the propagation of a merger shock in a clumpy ICM with self-consistent magnetic field structures. We computed the total integrated spectrum for the entire simulated relic across the same range of frequency of the real Toothbrush observations, as well as integrated spectra for three different sub-regions of the simulation, as shown in the inset of Fig.\,\ref{relic_denis}.

Note that these regions show very different dynamical properties and magnetic field topologies as detailed in Sec.\,3.6 of \citet{2019arXiv190911329W}. We computed the spectral slope of the emission for each different patch (see inset in Fig. \ref{relic_denis}) with a simple least square fit, assuming a $10\%$ uncertainty for flux below $1.4$\,GHz and $4\,\%$ at $1.4$\,GHz and above, similar to the real observation.

The simulation shows an intriguing similar spectral index, i.e., $\sim -1.16 \pm 0.02$ (see Fig. \ref{toothbrush_spectral_index}). This is true for the separate regions of the simulated relic, as well as for the whole relic. Like observations, it is remarkable that the spectra in the simulations are so similar in the different regions, despite the $\sim 1.2$\,Mpc extension of the relic and its clumpy morphology. This is at variance with what one might expect from the significant variations of Mach number at the shock front typically found in simulations \citep[][]{Hoeft2011, Skillman2013, 2018ApJ...857...26H, 2019arXiv190911329W}.  The combined effects of local variations in Mach number, 3D magnetic field fluctuations along the line of sight, the curved surfaces of realistic shocks, and the finite extension of the downstream cooling region of electrons can indeed conspire to broaden the individual spectral contributions and to converge on a very narrow range of power laws. In addition, it seems that the tail of the Mach number distribution, see e.g. Fig.~4 in \citet{2019arXiv190911329W} and Fig.~3 in \citet{Hoeft2011}, determines the average spectral index. Of course, while this first comparison can qualitatively explain the very narrow distribution of spectral indices in the Toothbrush relic, a larger statistics of simulations and improved aging models for relativistic electrons are needed in order to fully explain the surprisingly straight spectrum of the Toothbrush.

   
\subsection{Spectral index and curvature}

We examine the spectral index and curvature across the relics and compare them with basic models for cosmic-ray electron (CRe) injection and cooling. The difficulty in such spectral studies is that, in general, the curvature in the spectrum is very gradual, thus requiring sensitive and high-fidelity observations over a large frequency range at matched spatial resolution. The large size and high surface brightness of the Toothbrush allow us to conduct this type of analysis.

We created spectral index maps at the highest possible common resolution at all frequencies, namely at a resolution of $5\farcs5$. The same maps are used to create radio color-color diagrams. The surface brightness distribution of the relics shows quite sharp features, e.g., the outer edge presumably associated with the shock front and the steep downstream profiles at high observing frequencies. The high resolution also allows us to minimize the mixing of emission with different spectral properties within a single beam. For imaging, we used a common lower $uv$-cut and a uniform weighting scheme, see Sec.\,\ref{relic_flux} for details. Due to the different $uv$-coverages of the LOFAR, uGMRT and VLA data, the resulting images had slightly different resolutions, and therefore we convolved them to a common $5\farcs5\times5\farcs5$ resolution. For the spectral index maps at two frequencies, we considered only pixels with a flux density above 4.5$\,\sigma_{{\rm{ rms}}}$.

The resulting spectral index maps for several pairs of frequencies are shown in Fig.\,\ref{toothbrush_spectral_index}. Consistent with earlier works \citet{vanWeeren2012a,vanWeeren2016,Rajpurohit2018}, the spectral index maps of the Toothbrush show a spectral steepening, i.e. a spectral index patterns from 150\,MHz to 6\,GHz, which is remarkably homogeneous along the entire relic.

The spectral index at the outer edge of the relic, where the acceleration process is presumed to happen, is often used as an estimate for the ``injection'' index at the shock. We note that this should be done with caution if the resolution in the map and the downstream width at one of the relevant observation frequencies are similar, see \citet{Rajpurohit2018} for a discussion. For instance, the flattest spectral index of B2 and B3 towards the upstream region in the 150-650\,MHz map is apparently a bit flatter than in the 1500-3000\,MHz map. This could be caused by the fact that width in the 1500-3000\,MHz map the width of the relic is close the beam size. The 3000-6000\,MHz does not follow the trend of getting steeper, however, this frequency range suffers from a much lower signal to noise ration in B2 and B3 region. We would like to emphasize, at low frequencies, i.e., 150-650\,MHz, the downstream profile is wider, hence, this map is better suited to estimate the injection spectral index.

At the outer edge in the 150{-}650\,MHz map, we measure spectral indices in the range from $-0.65$ to $-0.80$. The variations in the measured spectral index at the outer edge along the relic may reveal larger variations in the injection spectral index than evident from the integrated spectra over larger regions. The variations may reflect inhomogeneities in the ICM, however, projection and smoothing effects may cause similar variations.

The spectra of radio sources are often characterized by simple semi-analytic models which take into account injection with power-law CRe momentum distribution and subsequent radiative losses. In the most basic scenario, the injection takes place at a single moment in time. Afterwards, the CRe momentum distribution ages according to synchrotron, Inverse Compton, and adiabatic losses. If the time scale at which the pitch angles between the electron momenta and the magnetic fields get isotropized is long, CRe with different pitch angles have different cooling times, therefore, the average of the spectra needs to be taken  \citep[`KP',][]{Kardashev1962,Pacholczyk1970}. If the pitch angles get isotropized on time scales much shorter than the relevant cooling times, all electrons cool with an average loss rate \citep[`JP',][]{Jaffe1973}. The latter scenario seems to be more appropriate for radio relics, because small scale magnetic field fluctuations are very likely present, possibly induced at the collisionless shock fronts \citep[e.g.][]{2013MNRAS.436..294B,2016MNRAS.462.2014D,2019arXiv190911329W}. More complex models considered here are therefore based on the JP-model. An obvious extension is to assume that the injection takes place not only at a single moment in time, but for an extended period. The injection may have stated at some time in the past and lasts until present \citep[continuous injection `CI', ][]{Pacholczyk1970} or may have stopped before present \citep[`KGJP',][]{Komissarov1994}.

Each model gives the CRe momentum distribution as a function of time after the injection started. The spectrum at any time may be characterized by a point in a color-color diagram \citep{Katz1993}. The time evolution results in a ``trajectory'' in the color-color diagram.  The four scenarios introduced above show distinct trajectories, see, e.g., \citet{Katz1993,Rudnick1994,vanWeeren2012a,Stroe2013}. It is worth noting that the trajectory in the color-color diagram is independent of the actual magnetic field.

\begin{figure*}[!thbp]
\centering
\includegraphics[width=0.49\textwidth]{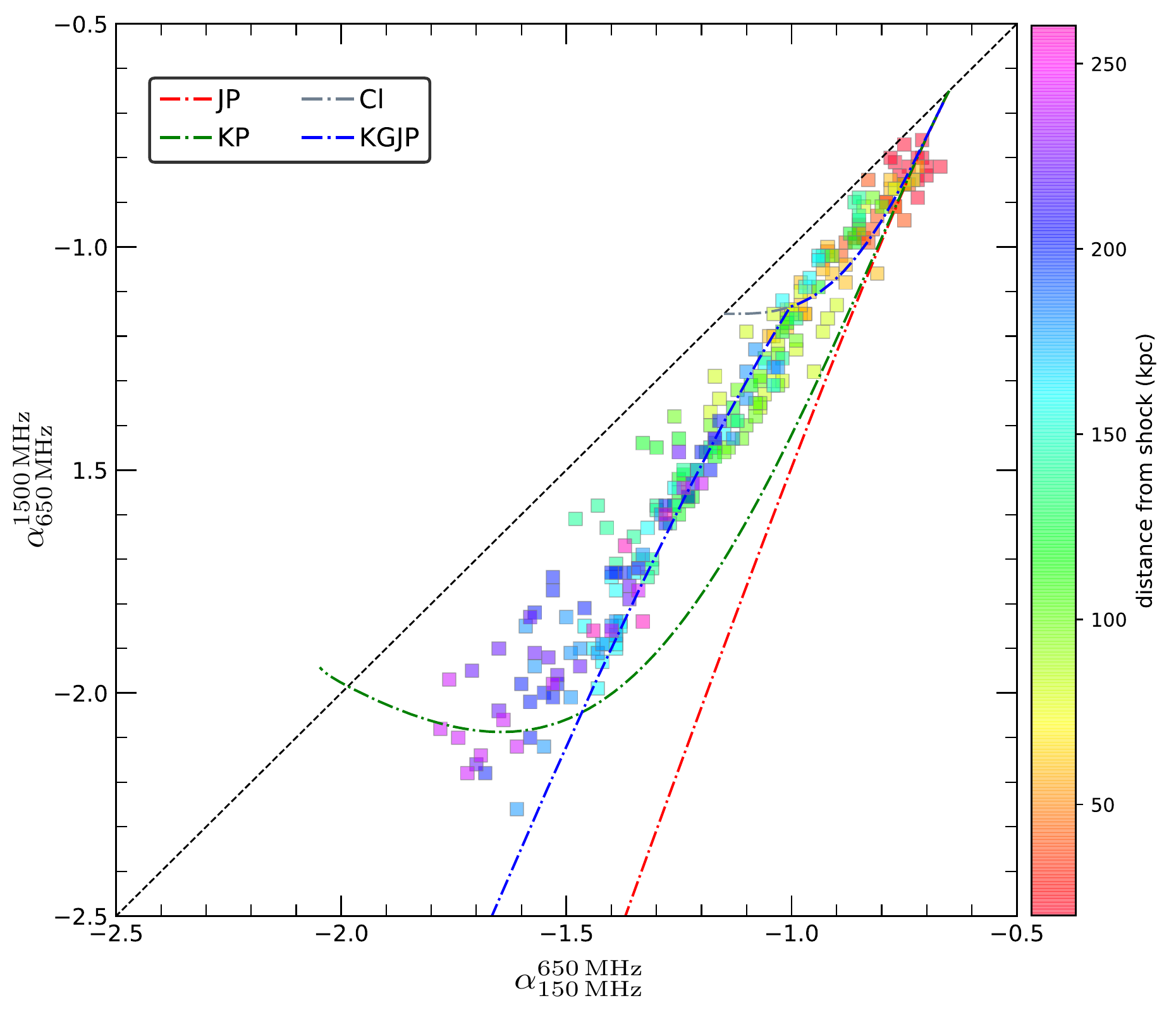}
\includegraphics[width=0.458\textwidth]{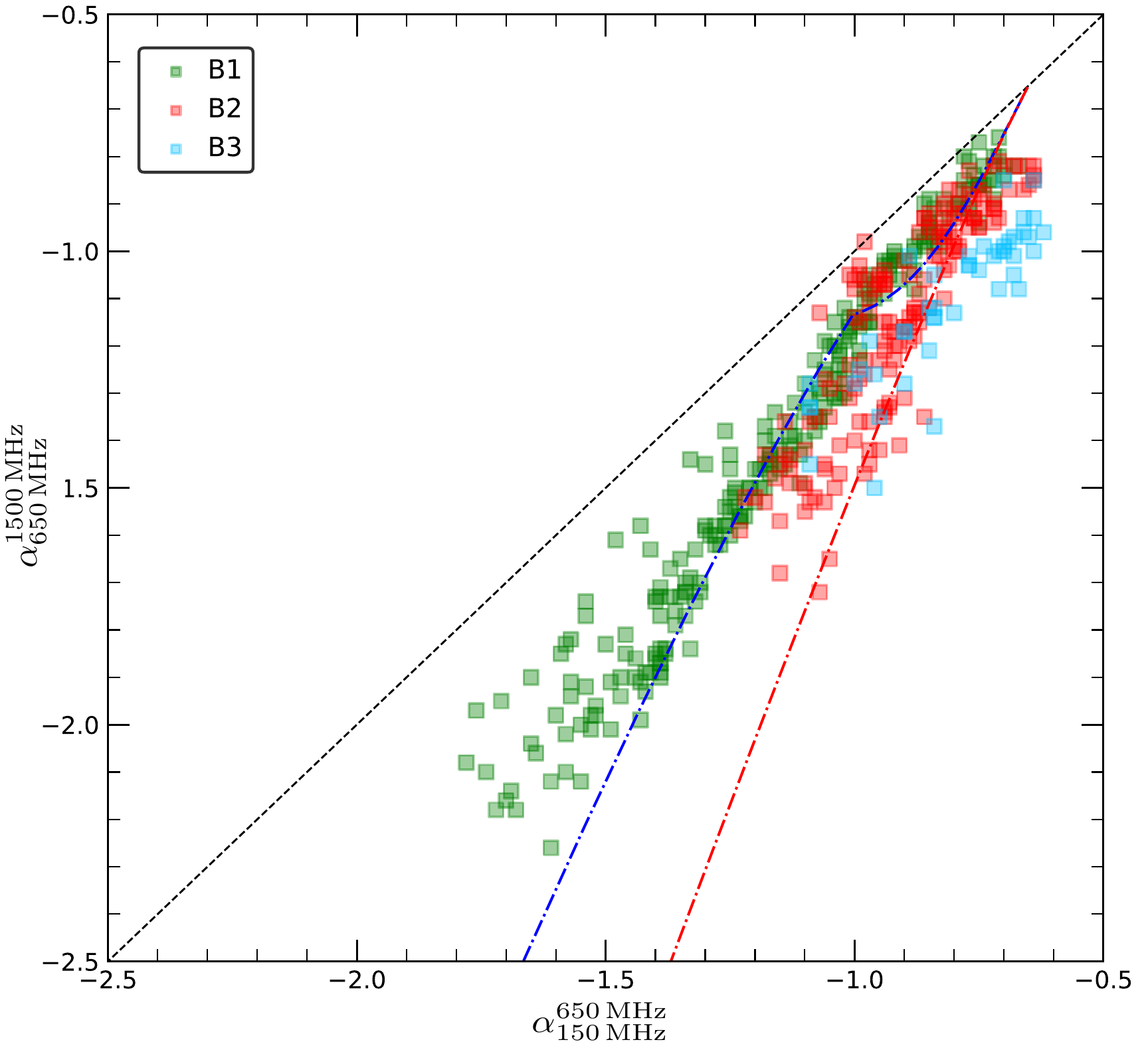}
\vspace{-0.2cm}
\caption{ \textit{Right}: Color-color diagram of the Toothbrush B1 region made at 5\farcs5 resolution and superimposed with the JP (red), KP (green), CI (gray), and KGJP (blue) spectral aging models obtained with $\alpha_{\rm inj}=-0.65$. The data match reasonably well the KGJP-model indicating that the spectral behavior can be understood assuming a propagating shock front which is slightly inclined to the line of sight. The color bar shows the distance from the shock front in kpc. Bottom:
 \textit{Left}: Color-color diagram of the B2 and B3 regions compared to the B1 region. The spectra of the B2 and B3 regions are apparently more curved than the ones of the B1 region. The KGJP curves are for particles injected continuously for about $0.3\times10^{8}$\,yr and $B=5\upmu\,\rm G$. To extract spectral index values, we create squared boxes with a width of $5\farcs5$, corresponding to a physical size of about 20\,kpc. Spectral index values were extracted from maps created using the IM6, IM11, IM15, and IM19 (see Table\,\ref{Tabel:imaging} for image properties).}
\label{ccplot_relic_all}
\end{figure*}

When the plasma, while aging, moves away from the location of injection, the spectral shape depends on the distance to the point of injection. In such a situation, the trajectory in the color-color diagram is obtained by depicting the color-color values of all distances. Even if actual injection cannot be resolved spatially, the extrapolation of the trajectory in the color-color diagram to the line of equal high-frequency and low-frequency spectral indices, i.e., the line of power laws, allows us to estimate the injection spectral index \citep{Katz1993,Rudnick1994}.

It is interesting to interpret the morphology of radio relics in terms of the spectral models described above. The spectral profiles of a CRe population injected at a planar shock front with a uniform Mach number and observed perfectly edge-on, is described by the KP and JP models. Since the plasma downstream to the shock moves away from the front, a spectrum of single spectral age is observed for each line of sight. If the shock front is inclined with respect to the line of sight, different spectral ages are present along the line of sight and contribute to the observed spectrum. If the line of sight goes through the shock plane the CI model provides the appropriate description. If the line of sight is far downstream that and does not intersect the shock front, the KGJP model provides the appropriate description.

\cite{vanWeeren2012a} studied the color-color diagram of the Toothbrush using three frequencies, namely 610\,MHz, 1.3\,GHz, and 2.3\,GHz. They found that the KGJP model provides a good fit for the entire Toothbrush. Here, we extend the color-color analysis using higher resolution maps and lower frequency data.  We do not use the S-band and C-band data because the width of the Toothbrush decreases at high frequencies; this would significantly reduce the area in which we could study the curvature. We use our spectral index map created between 150\,MHz and 610\,MHz and the one between 610\,MHz and 1.5\,GHz, see Fig.\,\ref{toothbrush_spectral_index}, for low- and high-requency spectral index maps, respectively.

The left panel of Fig.\,\ref{ccplot_relic_all} shows the trajectory in the color-color diagram for the region B1 and a comparison with the models introduced above. Similar to the findings by \cite{vanWeeren2012a} for the entire Toothbrush, the color-color distribution of the B1 region is basically consistent with the KGJP model but obviously inconsistent with all other models. This is not surprising, since it has been argued already, that the Toothbrush requires significant projection to explain the large downstream extent \citep{Rajpurohit2018}. Interestingly, at a distance of about 50 to 150\,kpc the trajectory in the color-color diagram of B1 shows some evidence for the kink, which is typical for the KGJP model. It marks the transitions from the line of sight intersecting the shock front (CI model) to the line of sight being further downstream (KGJP model). However, the trajectory in the color-color diagram after the `kink' is evidently flatter than the actual KGJP model. This might be caused by smoothing due to the resolution in the image and additional projection (the shock front does not have a sharp edge, it is more likely a projection of an approximately spherical profile). Overall, the KGJP model seems to provide an appropriate approximation, highlighting again that properties of the B1 region are significantly affected by projection.

We compare the trajectories in the color-color diagram for B2 and B3 to the one of B1, see the right panel of Fig.\,\ref{ccplot_relic_all}. Apparently, there is an offset between the curves for B1, B2, and B3.  This cannot be explained only by differences in the magnetic field, which simply shift points in the color-color diagram along the same locus. We speculate that this is caused by less projection, i.e., the shock front in the B2 and B3 region is less inclined w.r.t. the line of sight, hence the trajectory gets closer to the JP model here. Moreover, the interplay between the profile and the smoothing may plays a critical role since a more edge-on view would make the profiles narrower and the spectral indices more vulnerable to the smoothing effects discussed above. Therefore, a shock front which is better aligned with the line of sight may explain that the trajectories of the B2 and B3 region look different from B1, even though the overall spectrum is remarkably similar for all three regions.

Overall, the trajectories in the color-color diagram and the reasonable match with the KGJP model nicely support that the spectral indices and curvature found for the Toothbrush can be understood if assuming a shock front which moves outwards and is slightly inclined to the line of sight. A perfect match between the simplistic model and the observations cannot be expected since there are many effects which are not considered in the model. Considering the high sensitivity and resolution of the data we used, this finding shall be considered as a general caveat to radio observations, concerning the realistic limitations of model fitting to observed color-color spectra. 

\begin{figure}[!thbp]
\centering
\includegraphics[width=0.49\textwidth]{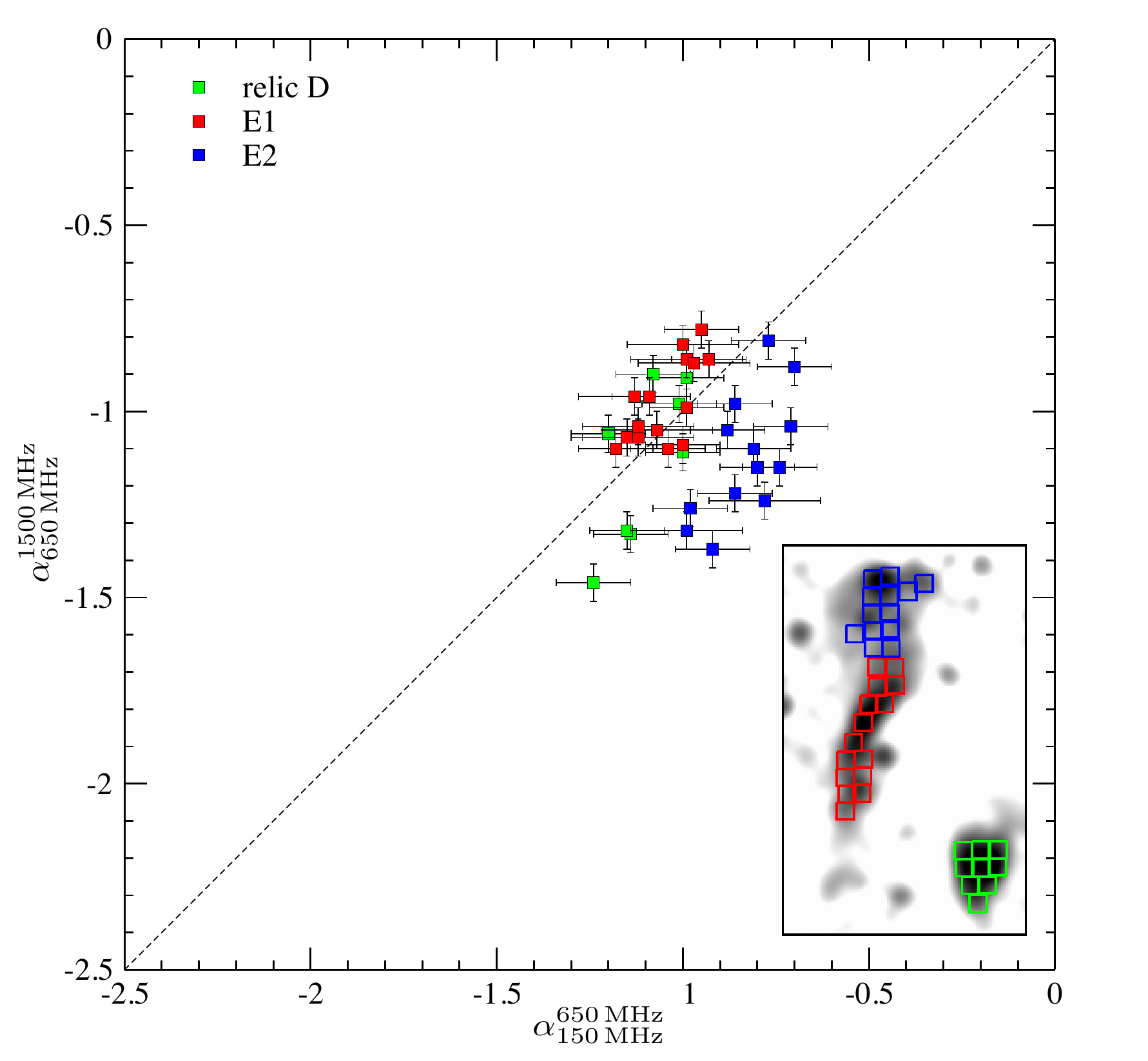}
\vspace{-0.6cm}
\caption{Color-color diagram of relics E and D. It is evident from the color-color plot that E1 and E2 are two different regions of the relic E with different ongoing processes. Spectral index values were extracted from maps created using the IM10, IM13, IM18, and IM21 (see Table\,\ref{Tabel:imaging} for image properties). The regions have a width of $16\arcsec$ corresponding to a physical size of about 60\,kpc. }
\label{ccplot_relic_all1}
\end{figure}

The spectral index maps of relics E and D do not show a strong spectral index gradient as a function of position, see Fig.\,\ref{halo_index}. To investigate whether they show any sign of spectral curvature, like the Toothbrush, we created a color-color diagram, using the spectral index extracted from $15\farcs7$ (i.e., the beam size) boxes covering the relic E and relic D, see the right panel of Fig.\,\ref{ccplot_relic_all1}. The regions are shown in the inset of the bottom right corner.  

We find that there are two different regions in the relic E. The points extracted across the E1 (red) part fall along the power-law ($\alpha^{{650\rm\,MHz}} _{{150\rm\,MHz}}=\alpha^{{3\rm\,GHz}}_{{1.5\rm\,GHz}}$) line, suggesting that at these locations  there is a very broad spectrum (a spectrum with a wide range of slopes at different locations), approaching a power law. On the other hand, for the E2 region (blue) the spectra are curved, i.e., the points below the power law line.  Such a curved spectra could be due to aging. From the color-color it is evident that E1 and E2 are different regions, with different processes going on. It remains an open question why the relic E shows such a distribution.   

It is interesting that the relics in 1RXSJ0603.3+4214 show different spectral properties. While the Toothbrush shows a strong spectral index gradient, no strong spectral gradient is seen across the relic E.  The integrated spectrum of the Toothbrush follows a single power law but the relic E instead shows a spectral break at high frequencies.


\subsection{Resolving discrepancies in the radio derived Mach numbers}
\label{sec::mach}

For a stationary shock in the ICM, i.e. the electron cooling time is much shorter than the time scale on which the shock strengths or geometry changes, and for a constant downstream magnetic field strength the integrated spectrum ($\alpha_{\rm int}$) is by 0.5 steeper than the injection spectrum ($\alpha_{\rm inj}$):
\begin{equation}
   \alpha_{\rm int}  
   = 
   \alpha_{\rm inj} - 0.5 \:\:
   .
   \label{approx}
\end{equation}

It is worth noting that the integrated spectrum is the sum of regions with very different CRe population ages, i.e., very different times have passed since the shock propagated over the emitting region. The oldest CRe populations contribute to the spectrum at the lowest frequencies. We find from the KGJP-model that a spectral index of $\alpha_{150\,\rm MHz}^{650\,\rm MHz} = -1.7 $ is reached at an age of roughly about 300\,Myr. The actual value depends on the assumed magnetic field strength. If the shock front propagates with a speed of $1000\,\rm km \, s^{-1}$, this corresponds to a distance of about 300\,kpc. The shock strength may change while passing this distance, hence, it is conceivable that the assumption of a stationary shock is not valid.

The spectral index of the outermost edge of the spectral index maps is considered to be close to the actual injection spectrum $\alpha_{\rm inj}$. We note that the combination of cooling distance, projection and smoothing in the radio image may make it impossible to measure the actual injection spectrum, however, high resolution maps at low observing frequencies may provide a reasonable estimate in this case. From our high-resolution spectral index maps between 150\,MHz and 650\,MHz, we find that the flattest spectral index at the outer edge is about $-0.65$. This values agrees well with the slope of the integrated spectrum, namely -1.16, and the assumption of a stationary shock. Our observations do not provide any evidence that, for example, a gradient of the shock strength affects the integrated spectrum although the low-frequency flux densities presumably have contributions from CRe populations that are rather old. On the contrary, our Toothbrush observations affirm that assuming a stationary shock is appropriate for radio relics.

Moreover, the color-color diagrams may provide a reliable estimate for the injection spectral index, see, e.g., discussion in \citep{Gennaro2018}. A KGJP model with an injection spectral index of $\alpha_{\rm inj}=-0.65$ provided a reasonable fit to the data of the bright B1 and B2 regions. Spectral shapes in the B3 region are presumably affected by subtle effects. Therefore, the color-color analysis again supports that the relation between the injection and integrated spectrum can be described assuming a stationary shock.

If the spectral index of the integrated spectrum $\alpha_{\rm int}=-1.16\pm0.02$ reflects the strength of the shock, we obtain a Mach number of  $3.7^{+0.3}_{-0.2}$. This is also consistent with our analysis of the surface brightness profiles in \citet{Rajpurohit2018}. For the Toothbrush all methods for deriving the Mach number from the radio data provide a consistent value. The derived shock strength is evidently much higher than obtained from the X-ray data \citep{vanWeeren2016}. We also note that the analysis of the simulation, see Sec.\,\ref{simulation_part}, indicates that the shock front actually shows a distribution of Mach numbers, the single Mach number derived above can only roughly characterize the shock.


\subsection{Radio luminosity vs. thermal energy content}
\label{efficiency}

The brush (B1) of the Toothbrush is very radio bright. It has been argued that the high luminosity of the Toothbrush --and of other radio relics as well-- is only possible when a significant fraction of the kinetic energy flux through the shock front is channelled into the acceleration of relativistic electrons (CRe) \citep{vanWeeren2016, 2019arXiv190700966B}. If the CRe originate from diffusive shock acceleration (DSA) of thermal electrons this would require a very high acceleration efficiency; in case of weak magnetic fields and a weak shock, the luminosity can not be explained by the standard DSA scenario. The tension between the relic luminosity and the kinetic energy flux might be lowered assuming an upstream population of mildly relativistic electrons which requires less energy input to produce the observed synchrotron luminosity. 

Since we have measured the spectrum of the Toothbrush over a wide frequency range, we can address this problem adopting a slightly different viewpoint and comparing the radio luminosity of the emitting volume to the thermal energy content of that volume {\it without} considering any particular acceleration mechanism. 

A small ICM mass element downstream to the shock may emit at the frequency $\nu$ with the emissivity  $\epsilon_{\nu}(t)$. Due to the radiative losses of the relativistic electrons the emissivity decreases with distance to the shock front. We simplify the profile and assume that the mass element emits with a constant emissivity $\epsilon_{\nu}$, from the time that it has  just passed through the shock front to the time $t_{\rm d}(\nu)$ . The  time $t_{\rm d}(\nu)$, after which the emissivity drops to zero, is related to the frequency dependent downstream extent of the emitting volume by $d(\nu)= t_{\rm d}(\nu) \cdot v_{\rm down}$, where $v_{\rm down}$ denotes the relative speed of the downstream material with respect to the shock front. The luminosity of the entire emitting volume is given by  $P_\nu =  \epsilon_{\nu} \cdot A \, d(\nu)$, where $A$ is the surface area of the shock front. 

The synchrotron emission over all frequencies of a volume $\Delta V$ during the `luminous period' $t_{\rm d}(\nu)$ is given by the integral 
\begin{equation}
    E_{{\rm sync}}
    =
    \Delta V \, \int   \epsilon_{\nu} \: t_{\rm d}(\nu) \: {\rm d} \nu 
    =
    \frac{\Delta V}{A\,v_{\rm down}} \: \int   P_\nu \: {\rm d} \nu  
    \:.
\end{equation}
The ratio of the synchrotron emission to the thermal energy in the volume, $E_{\rm th}= \Delta V \, n \, \overline{e}$, amounts to
\begin{equation}
    \frac{E_{\rm sync}}{E_{\rm th}}
    =
    \frac{\int   P_\nu \: {\rm d} \nu }{ A \, v_{\rm down} \, n  \, \overline{e} }
    \: ,
    \label{eq:sync-th-ratio}
\end{equation}
where $n$ denotes the particle density and $\overline{e}$ the average thermal energy per particle. The rest-frame luminosity is related to the measured flux density $S_\nu$ by
\begin{equation}
    P_\nu = 4 \pi \, D_{\rm L}^2(z) \, (1+z)^{-(1+\alpha_{\rm int})} \, S_\nu 
    \: ,
\end{equation}
where $D_{\rm L}$ denotes the luminosity distance. 

We assume that the spectrum follows a power law in the observed frequency range with 210\,mJy at 1.5\,GHz and a spectral slope $\alpha_{\rm int}=-1.16$. Furthermore, we adopt that a particle density in the downstream region of $10^{-4}\, {\rm cm^{-3}}$, an average energy per particle of 6\,keV, and a shock surface area of $(0.7 \, {\rm Mpc})^2$. 

Limiting the frequency integral in Equation~\ref{eq:sync-th-ratio} to only the observed range from 150\,MHz to 8\,GHz, we find a ratio $E_{\rm sync}/E_{\rm th} = 5 \times10^{-3}$.   The spectrum of the brush shows a perfect power law, with no indications for  bending at low or high frequencies, see Fig.\,\ref{spectrum_cc}. Thus, according to the theory of synchrotron emissivity of monochromatic electrons \citep[][]{1982AN....303..142R}, the observed power law shape of the emission spectrum also implies radio emission outside of the frequency range of the observations.

To estimate the minimum luminosity outside the observed frequency range we model the electron spectrum by assuming a constant momentum distribution up to a Lorentz factor $\gamma_{\rm knee}$, a power law momentum distribution from $\gamma_{\rm knee}$ to $\gamma_{\rm max}$, and no electrons with Lorentz factor above $\gamma_{\rm max}$. The shape is intended to mimic the distributions derived in \citet{1999ApJ...520..529S}. Assuming a magnetic field strength of $1\,\upmu{\rm G}$, we find that $\gamma_{\rm knee}$ has to be <3000 and $\gamma_{\rm max}$>$10^5$ to avoid that a clear bending of the spectrum within the observed frequency range becomes noticeable. This implies that at least the same amount of energy must be radiated via synchrotron emission outside of the observed frequency range. Therefore, the observed power law spectrum of the brush region implies that synchrotron radiation is equivalent to a loss of at least 1\,\% of the thermal energy.

It should be noted that this result depends neither on the actual acceleration process nor on the spectral energy distribution of the electrons. Relativistic electrons will also emit via the inverse Compton (IC) process. The ratio of synchrotron to IC losses is determined by the ratio of the energy densities of magnetic fields, $u_B=B^2/8\pi$, and Cosmic Microwave Background photons. The equivalent magnetic field of the latter is given by $B_{\rm CMB} = 3.24\,(1+z)^2\,{\rm \upmu G}$. With a magnetic field $B$ below $ B_{\rm CMB}$ a significant fraction of the energy content in a volume $\Delta V$ would be emitted via IC emission. This would clearly affect the thermal evolution of the intra-cluster medium and would have an impact on the ICM evolution. Since such an effect has not been observed, low magnetic fields in the emission region of the relic can be excluded. This is in agreement with the non-detection of IC emission in hard X-ray. The analysis of recent Suzaku observations provides a lower limit for the magnetic field strength, namely $ 1.6 \, \rm \upmu G$ \citep[][]{Itahana2015}.


\section{Halo structure and spectrum analysis} 
\label{halo}

The cluster shows low surface brightness radio emission with a morphology remarkably similar to the ICM distribution \citep{vanWeeren2016, Rajpurohit2018}.  Even if it is widely accepted that halos are related to ICM turbulence, details of their origins  are not well understood. Our deep observations allow us to study the spectral properties of the radio halo over a wide frequency range and to assess the physical mechanisms \citep[e.g.][]{Brunetti2001} powering the radio emission.

\subsection{Integrated spectrum of Halo C}
\label{halo_spectra}

In order to obtain the integrated radio spectra of the halo, we measure the flux densities from the $15\arcsec$ radio maps, where the signal-to-noise ratio is high enough. To ensure that we recover flux on the same spatial scales for all frequncies, we again produced images with a common lower uv-cut of $0.4\,\rm k\uplambda$.  The regions where the flux density is measured are indicated in the left panel of Fig.\,\ref{region}. Since the halo has a very low surface brightness, we average for each observation the entire bandwidth, as given in Table\,\ref{Tabel:Tabel4}. The resulting integrated spectrum is shown in Fig.\,\ref{region}. 

 Remarkably, the halo shows a power law with no indications of a spectral steeping, similar to the Toothbrush relic, when we restrict the analysis to the main halo region without relic+halo and region S. The integrated radio halo emission in that region between 150\,MHz and 3\,GHz, has a spectral index of $\alpha_{\rm halo}=-1.16\pm0.04$, which is, even more surprisingly, identical to the integrated relic spectral index. 
This spectral index of the halo is in agreement with earlier studies \citep{vanWeeren2016, Rajpurohit2018}. However, our results show for the first time that the halo follows indeed a power law over a wide frequency range. Such a wide frequency range spectral coverage is only available for a few other clusters \citep[for a review see][]{Feretti2012}. Those reveal a large variety of integrated spectral indices and of spectral shapes. For instance, \citet{Thierbach2003} reported for the halo in the Coma cluster a significant steepening above about 1.5\,GHz from about -1.2 to -2.3. We note that \cite{Xie2020} recently reported a curved spectrum for the radio halo in Abell\,S1063 and Abell\,370. In contrast, \citet{2000ApJ...544..686L} and \citet{Shimwell2014} found for the Bullet cluster a rather constant slope of -1.3.  Interestingly, for the majority of halos with flux density measurements at more than two frequencies, as reported in \citet[][]{Feretti2012}, the integrated spectrum is either steep ($\lesssim -1.5$) or shows a steepening. Either the halo in 1RXS J0603.3+4214 is an unusual one, or the superb data quality permits us to do a much better job removing contaminating sources and distinct emission regions, such as region S and the relic-halo overlap region, as discussed in the next section.

\begin{figure*}[!thbp]
    \centering
    \includegraphics[width=1.0\textwidth]{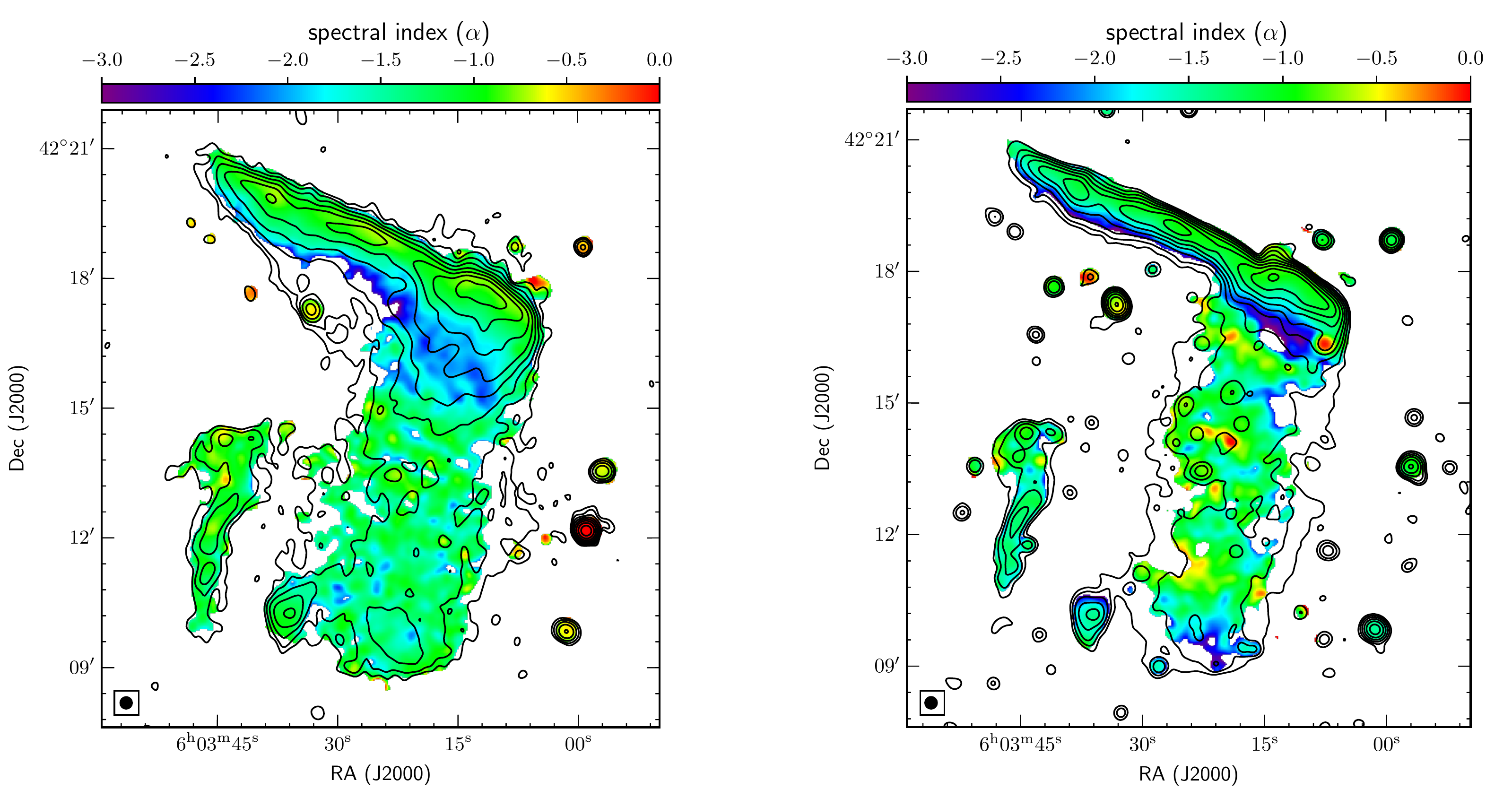}
  \vspace{-0.5cm}
    \caption{Spectral index map of the radio halo in 1RXS\,J0603.3+4214 at $15\farcs7$ resolution. \textit{Left}: Low frequency (150 and 650\,MHz) spectral index map. Contour levels are from the LOFAR and drawn at $\sqrt{[1, 2,4, 8,\dots]}\,\times\,4\,\sigma_{{\rm{ rms}}}$. \textit{Right}: High frequency (1.5 and 3\,GHz) spectral index map. Contours are from the VLA L-band and are drawn as in the left panel. These maps confirm a remarkable uniform spectral index distribution across the halo C between 150\,MHz to 3\,GHz. It is evident that the relic+halo region shows spectral steeping at low frequencies while there is a uniform spectral index distribution at high frequencies. These maps were created using the IM10, IM13, IM18, and IM21 (see Table\,\ref{Tabel:imaging} for image properties). The color bar shows spectral index $\alpha$.}
   \label{halo_index}
\end{figure*}


\subsection{Impact of SZ on the halo flux measurements?}

To obtain a reliable radio spectrum for the halo, we need to consider possible corrections for the Sunyaev-Zeldovich (SZ) effect. The SZ effect is the distortion of the blackbody spectrum of the cosmic microwave background (CMB) due to inverse Compton scattering of CMB photons by free relativistic electrons in the ICM.  As a result, the SZ effect can cause a {\it steepening} in the observed spectrum at $\nu>1\rm\,GHz$, by lowering the zero point of the radio emission as a function of frequency \citep[e.g][]{2002A&A...396L..17E,2013A&A...558A..52B,2014MNRAS.441L..41S,2016A&A...591A.142B}.  Note that although the uniform CMB is not visible to the interferometers, small scale variations, such as those due to the SZ effect, will be detectable.

To check whether this is the case for our observation of the halo, we computed the expected level of the SZ decrement for the halo in 1RXS J0603.3+4214 with a realistic model of its gas density. We started by simulating a gas density and temperature distribution for a simple $\beta-$model, which we perturbed by introducing elongation along the direction connecting to the radio relic using an aspect ratio of $1$:$0.8$, as well as a second massive sub-clump (with a total mass $M_{\rm 100} \approx 10^{14} M_{\odot}$), also described by a simple $\beta$-model and located at the approximate position of the X-ray substructure in 1RXS J0603.3+4214. 

When integrating within a region corresponding to the halo C, we estimate a total SZ decrement at $3$\,GHz of $-2.97$\,mJy.  However, the total  observed SZ decrement from the halo region is further reduced to $-1.12$ mJy, if we apply to the simulation the same $\geq 0.4$ $\rm k \uplambda$ uv-cut as in the observation. We investigated variations of this baseline model, i.e., by changing the mass of the second clump and/or its distance along the line of sight; these do not change our result. We thus conclude that the SZ decrement can only be $\leq 6$ $\%$ of the radio flux density in the entire halo region, and therefore it does not significantly contaminate the measured emission.


\subsection{Spectral index and curvature distribution}
\label{halo_indexmaps}

We construct high and low frequency spectral index maps of the halo to investigate if there are any spatial variations across the radio halo. We use the images described in Sec.\,\ref{halo_spectra}. The low frequency spectral index map is created using the 150\,MHz and 650\,MHz and the high frequency one between 1.5\,GHz and 3\,GHz. We do not subtract the extended or point like discrete sources embedded in the halo. Pixels with flux densities below 4$\sigma_{{\rm{ rms}}}$ were blanked. We note that above 1\,GHz, there are around 32 faint point sources embedded in the halo region, while at 650 and 150 MHz we identified around 10 and 4 sources, respectively. This do not allow us to subtract discrete sources embedded in the halo region.

Fig.\,\ref{halo_index} shows the high and low frequency spectral index map of the halo. As visible from both the high and low frequency spectral index maps, the spectral index across the central part of the halo, excluding the relic+halo and region S, is more or less uniform. To investigate it further, we employ the color-color diagram. The regions where the spectral indices were extracted are marked with green in the left panel of Fig.\,\ref{region}. The size of the each box is $15\farcs7$, corresponding to physical sizes of about 57\,kpc. 

\begin{figure}[!thbp]
  \centering
\includegraphics[width=0.49\textwidth]{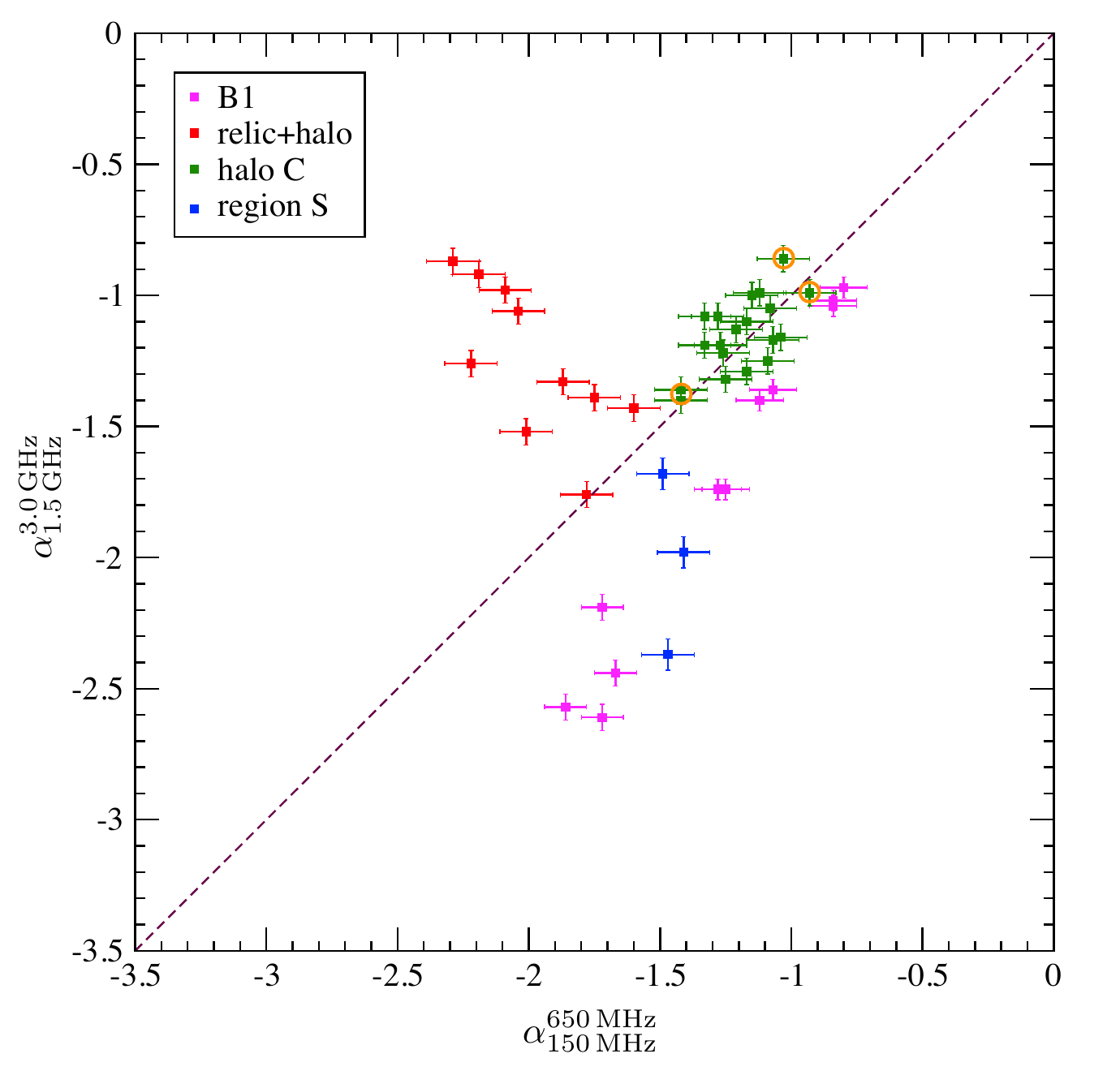}
  \vspace{-0.5cm}
   \caption{Radio color-color diagram of the halo in 1RXS\,J0603.3+4214, using the spectral indices extracted in $15\arcsec\times 15\arcsec$ boxes shown in the left panel of Fig. \ref{region}. The contaminated regions in the halo C are shown with orange circle.  The relic+halo (red) indicate combined emission from the B1 region of the Toothbrush and the halo. The halo C points are clustered around the power law line, indicating no sign of spectral curvature. The relic+halo points lies in the concave part of the color-color plot, suggesting superposition of the halo and the relic emission.}
   \label{halocc}
  \end{figure}

The resultant color-color distribution is shown in Fig.\,\ref{halocc}. The halo points (green) are indeed clustered near the black-dashed line. This is a clear signature of a power-law. It is evident from Fig.\,\ref{halocc} that the spectral index of the halo is uniform over a broad range of frequencies. The is no indication of spectral curvature between 150\,MHz and 3\,GHz, even in regions as small as the selected boxes.

The radio halo shows a quite homogeneous surface brightness which correlates with the X-ray surface brightness \citep{Rajpurohit2018}. As discussed above, the spectral index distribution is quite uniform and there is no indication of spectral curvature. The power-law spectrum over a large frequency range requires an electron momentum distribution which follows a power law as well at least for about one order of magnitude in momentum. In contrast to the relic, the power-law spectrum may arise from averaging along the line of sight through the ICM if the halo emission originates from volume filling turbulence. This would allow for local deviations from the average spectrum.

In the basic model explaining radio halos by turbulent re-acceleration of mildly relativist electrons, the acceleration is caused by electron scattering at magnetohydrodynamic waves which are excited as part of turbulent cascade \citep{2004MNRAS.350.1174B,Brunetti2014}.  We note that based on homogeneous conditions the simulated spectra rather produce curved spectra in the relevant electron momentum regime. A power-law emission spectrum, as the observed one, is not expected from the basic model. Indeed, \cite{Xie2020} have recently used the curved spectrum for the radio halos in Abell\,S1063 and Abell\,370 to support the re-acceleration scenario.

Adopting the critical frequency of synchrotron emission, we estimate that the Lorentz factor of the electrons emitting at the highest observed frequency for the halo, 3\,GHz, amounts to $3\cdot 10^4$ for a magnetic field of $1\,{\rm \upmu G}$. For these electrons the radiative loss time is as short as a few ten Myr \citep{1999ApJ...520..529S}, which is of the same order or somewhat below the expected acceleration time scale \citep[see e.g. Fig. 13 in][]{2004MNRAS.350.1174B}.

These timescales are much smaller compared to the crossing time of the halo region ($\sim 1 \rm ~Gyr$). This reinforces that the powering of radio emitting electrons by turbulent motions must happen throughout the radio halo volume, otherwise we would expect local variations in the curvature, even in projection. The acceleration timescale must also be significantly shorter than the cascade time along the Kolmogorov cascade, which also is of the order of the crossing time \citep[e.g.][]{Brunetti2014}.  Therefore, the observed uniformity of the spectrum suggest that the halo emission is being being powered by an acceleration mechanism which is uniformly spread across the halo region. Why the spectrum is homogeneously a power law remains to be understood.  

\subsection{Radio halo-relic connection}
\label{halorelic}
The halo and the "brush" region of the Toothbrush are connected by a region, denoted as "relic+halo" (for labeling see the left panel of Fig.\ref{region}). A comparable example of the morphological connection between radio relic and halo was observed in CIZA\,J2242.8+5301 where the elongated halo is morphologically connected to the northern and the southern relic  \citep{Hoang2017}. 

We examine the spectral properties of radio emission in the apparent overlap regions to see if they are produced by a unique electron population or rather by the superposition of two separate particle populations along the line of sight. This is key to disentangling the complex chain of (re)acceleration events in the ICM \citep{Hoang2017}. The spectral index across the relic+halo region, between 150 MHz and 1.5 GHz, gradually flattens (from -2 to -1.2)  over a distance of about 400\,kpc and returns to being uniform in the rest of the halo region \citep{vanWeeren2016,Rajpurohit2018}. It has been speculated that this spectral behavior indicates a possible connection between the shock and the  turbulence \citep{vanWeeren2016,Hoang2017}. 

The spectral index variations across the relic+halo region appear different at low and high frequency. At low frequencies, where the relic emission dominates that of the halo, the spectral index across the relic+halo gradually steepens, see the left panel Fig.\,\ref{halo_index}. Such a spectral steepening is expected in the downstream areas of relics when electrons undergo energy losses. 
In contrast, at high frequency, where the halo dominates, the relic+halo region shows a constant spectral index distribution like the rest of the halo, see the right panel Fig.\,\ref{halo_index}.

The above trends suggest that both halo and relic emission is present in the overlap region, in different proportions at different frequencies.  A clearer demonstration of this is seen in the color-color plot, where the relic+halo points (red points) lie above the power-law line. These points represent places in the spectrum with positive (convex) curvature, i.e., the spectrum is flatter at higher frequencies. This the expected behavior when there is a superposition of different emission spectra along the same line of sight. Although spectra with positive curvature can also be produced by a similarly-curved electron energy distribution, there is no need to invoke that here. 

We find that for the regions with concave curvature, the range of $\alpha^{{650\rm\,MHz}} _{{150\rm\,MHz}}$ is -2.30 to -1.50 and $\alpha^{{3\rm\,GHz}}_{{1.5\rm\,GHz}}$ is -1.80 to -0.86. It is consistent that the steeper (lower frequency) emission component is from locally shock accelerated aged electrons (i.e., relic region) and the flatter (higher frequency) is due to the turbulent re-acceleration of electrons.

The southern edge of the radio halo in 1RXS\,J0603.3+4214 coincides with a shock front detected in \textit{Chandra} observations\citep{vanWeeren2016}. In a handful of clusters the radio halo emission seems to be bounded by cluster shock fronts \citep{Markevitch2005,Brown2011,Shimwell2014}. The nature of these sharp edges is still unclear. Data points extracted across the region S (blue) lie below the power-law line. In the color-color plot, the region S appears different than the rest of the halo. The same region also appears distinct in the X-ray vs radio correlation \citep{Rajpurohit2018}, suggesting possible connection with the southern shock front, instead of a simple continuation of the halo.


\subsection{Modeling of relic and halo integrated spectrum}
\label{B1modelling}
We explore here the conditions leading to the integrated spectrum of relic B.  This involves understanding the aging of the electrons, in particular in the downstream region of B1,  where we have the best data. We model the aging of radio emitting particles under realistic yet simple conditions. We consider a simple "single-zone" model, assuming a single value of electron density, temperature and magnetic field across the entire B1 region. For simplicity, we also neglect the possible role of additional stochastic acceleration by cluster turbulence \citep[e.g][]{2014MNRAS.443.3564D} or the re-acceleration of fossil electrons in the shock downstream \citep[e.g.][]{ka12}. 

In our reference model, we fix the maximum length of the downstream cooling region, the downstream gas density and temperature to $l_{\rm d}=855 \rm ~kpc$, $n_{\rm d}=10^{-3} \rm cm^{-3}$ and $T_{\rm d}=9.7 \cdot 10^{7} \rm K$, respectively. For the maximum length  $l_{\rm d}$ we adopted the largest length of downstream emission visible in the LOFAR image. The downstream magnetic field is set to $B_{\rm d}=1.0 \rm\,\upmu G$ and the injection Mach number to $\mathcal{M}_{\rm inj}=3.7$. The energy flux that goes into the acceleration of electrons is given by 
\begin{equation}
  \Phi_e 
  =
  \frac{1}{2} \xi_{\rm e} c_{\rm s}^3
  \mathcal{M}^3_{\rm inj}
  \rho_{\rm p} \cdot A^2, 
\end{equation}
where $A$ is the assumed shock surface ($A \approx L^2$ where $L$ is the length of B1 region),  $c_{\rm s}$ and $n_{\rm p}$ are the pre-shock sound speed  and gas density, respectively. These values are computed based on the 
standard jump conditions for a shock strength $\mathcal{M}_{\rm inj}$. The acceleration efficiency of electrons at the shock is set to a fix reference value of $\xi_{\rm e} \approx \cdot 10^{-3}$, i.e., $0.1 \%$ of the ram kinetic energy of the shock goes into the acceleration of supra-thermal electrons. This value is in line with the value estimated  previously in Equation~\ref{eq:sync-th-ratio}, provided that $\xi_{\rm e}$ is referred to the shock kinetic energy flux while in Sec. \ref{efficiency} we analyzed the ratio between the energy of relativistic electrons and the thermal gas.

In a simple 1-dimensional shock propagation, the radiative lifetime of electrons along the downstream region is $t_{\rm rad} = x_{\rm d}/v_{\rm s}$, where $v_{\rm s}$ is the shock velocity and $x_{\rm d} \leq l_{\rm d}$ is the 1-dimensional coordinate in the downstream direction, which we assume is pointing towards the cluster center. The time-dependent diffusion-loss equation of CRe within each integration step in our algorithm \citep[e.g.][]{Kardashev1962,1999ApJ...520..529S} is solved with the \citet{1970JCoPh...6....1C} finite difference scheme. We use N=20,000 equal energy bins in the $10 \leq \gamma \leq 2 \cdot 10^5$  ($\gamma$ is the Lorentz factor of electrons) energy range and an adaptive time stepping algorithm with $0.01 \rm ~Myr \leq \Delta t \leq 100 \rm ~Myr$. The overall radio spectrum in the shock downstream is obtained by numerically integrating the synchrotron emission from the distribution of accelerated particles \citep[e.g.][]{1965ARA&A...3..297G}, applying the JP aging model for simplicity (the general conclusions here do not change if we adopt a different aging model, as in Sec.5.3 above).

\begin{figure}[!thbp]
    \centering
    \includegraphics[width=0.49\textwidth]{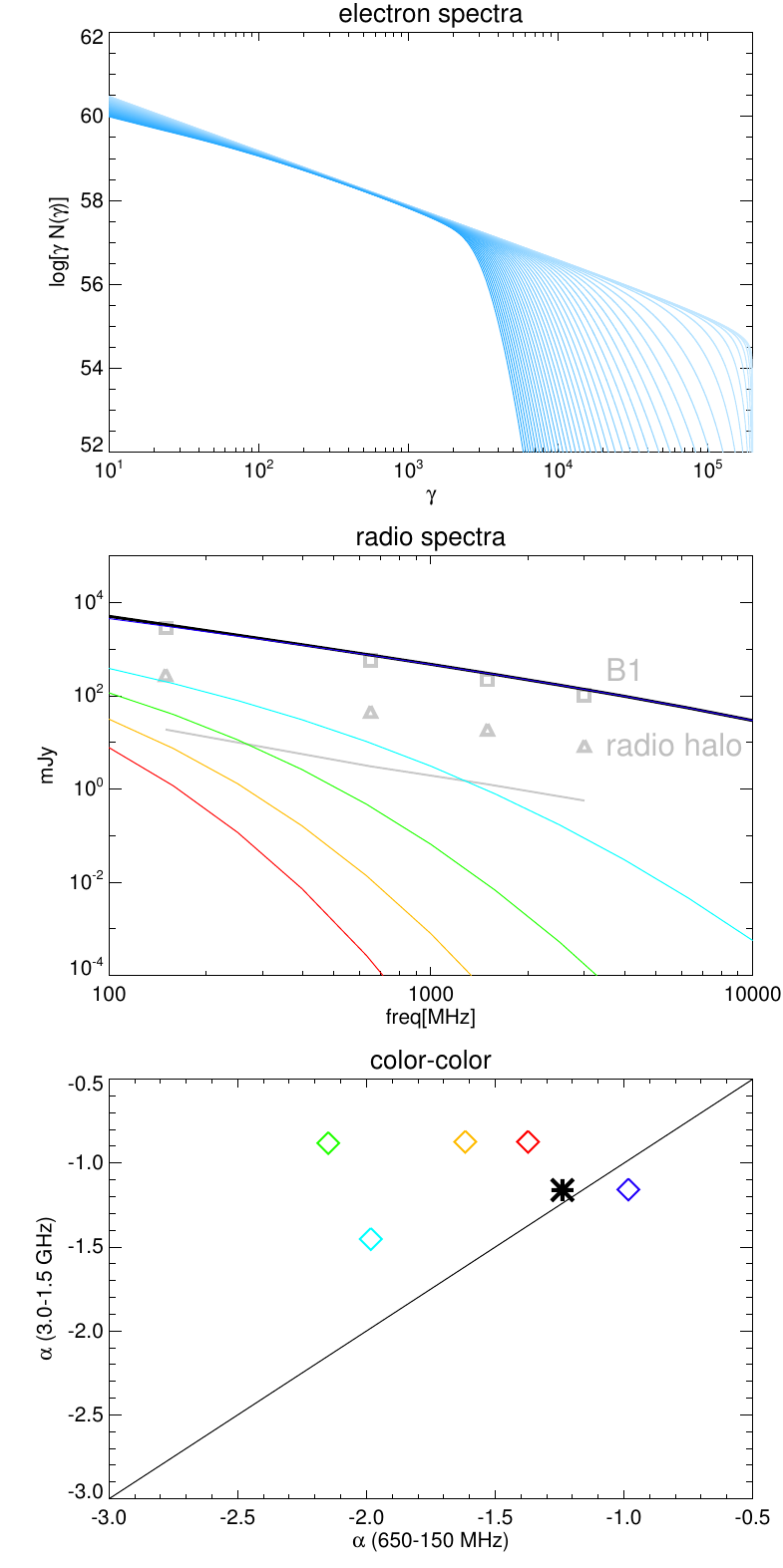}
  \vspace{-0.2cm}
    \caption{Top panel: evolving spectrum of electrons accelerated by a $\mathcal{M}= 3.8$ shock and aging in a uniform $2.0 ~\rm \upmu G$ magnetic field (see text for other model parameters). Middle panel:  evolving radio spectra as a function of time. The top black spectrum gives the integrated spectrum for the region corresponding to the entire B1 region, while the colored lines show the spectra within different $15\arcsec\approx 54 \rm ~kpc$ thick stripes perpendicular to the shock propagation axis. The additional gray symbols show the observed total integrated spectrum for B1 (top squares) and the radio halo (lower tringles), while the gray line gives the estimated contribution from the radio halo within each single $15\arcsec$ stripe (assuming a flat brightness distribution). Bottom panel: color-color plot obtained by comparing the  $\alpha^{{650\rm\,MHz}} _{{150\rm\,MHz}}$  and $\alpha^{{3\rm\,GHz}}_{{1.5\rm\,GHz}}$ spectral slopes within the stripes with corresponding colors, as in the central panel.}
      \label{fig_sim_color_color}
  \end{figure} 

The top panel in Fig. \ref{fig_sim_color_color} shows the time evolution of electrons injected at the shock and downstream cooling. The integrated spectra corresponding to the entire B1 and within different $15"$ width stripes is shown in the middle panel. First, we notice that to a good approximation our reference model is able to reproduce the observed B1 spectrum over a wide frequency range, which makes our modeling of the downstream aging of the spectrum a good enough assumption for the sake of the following steps. 

Secondly, to investigate the contamination of the B1 by the halo emission, we test a simple toy model scenario, in which we assume that the relic and the halo are not physically connected (hence their particle populations do not mix) but their emission is observed to be overlapping just in projection. We assume that the halo extends all the way up to the B1 region, at the same brightness that it is just below B1, and with an average spectral index. As a next step, we compute its contribution within each $15\arcsec$ stripes (as shown by the grey line in Fig.\ref{fig_sim_color_color}). This allows us to estimate the observable spectral curvature in the entire B1 region, by including both the radio components in an approximate way. 

The resultant color-color plot for the different regions, using $\alpha^{{650\rm\,MHz}} _{{150\rm\,MHz}}$  and $\alpha^{{3\rm\,GHz}}_{{1.5\rm\,GHz}}$ is shown in the bottom panel of Fig. \ref{fig_sim_color_color}. Our observations and modeling give a similar trend in the color-color plane for the B1 region contamination by the halo. The concave regions in Fig.\,\ref{halocc} and \ref{fig_sim_color_color} seem indeed compatible with the superposition of the increasingly steeper spectrum from the older particles accelerated by shock at B1, and the power-law spectrum of the halo. It is important to emphasize here that in  this simplistic model there is no ongoing particle acceleration as we go downstream of B1, because there is quite significant aging. On the other hand, we assume here a constant spectrum for the radio halo, on the 
basis that the energization of electrons may be happening in a very distributed and spatially uniform way \citep[e.g.][]{Brunetti2001,Brunetti2014}.

The concave curvature should, therefore, mark the frequency range (which decreases moving away from the B1 edge) at which the radio halo component dominates over the radio relic one, moving towards the center of 1RXS\,J0603.3+4214. 
According to this simple explanation
there is no need to assume that turbulence (or in general, any process responsible for the appearance of the radio halo) is acting on fossil radio electrons injected by the shock responsible for the B1 relic, also because in this case the low energy particle would be accelerated first, and the spectrum would be flatter at low frequencies, and steeper at higher \citep[e.g.][and references therein]{Brunetti2014}. 

Testing the hypothesis that the two emissions are related to physically disconnected acceleration regions is not easy, but in principle it can be done by detecting polarized emission from the intermediate relic+halo region of 1RXS\,J0603.3+4214, and using Rotation Measure  Synthesis \citep[e.g.][]{2005A&A...441.1217B} to detect a significant difference in Faraday depth of the two emitting structures. However, based on existing data (Rajpurohit et al., in prep.) no polarization is detected with available exposures in this region. 

\section{Conclusions and Summary}\label{sec::summary}

We  presented results from deep, wide-band uGMRT (550-750\,MHz) and VLA (2-8\,GHz) radio observations of the galaxy cluster 1RXS\,J0603.3+4214. We focused on the diffuse extended radio sources in the cluster, namely the Toothbrush relic, fainter relics E and D, and a large extended radio halo which basically follows the X-ray surface brightness distribution of the ICM. These observations were combined with the existing VLA L-band (1-2\,GHz) and LOFAR (120-180\,MHz) data \citep{vanWeeren2016,Rajpurohit2018} in order to carry out a detailed radio continuum and spectral study of the cluster. From these observations, we have been able to characterize the integrated spectrum of the Toothbrush relic over an unprecedented frequency range. Here, we summarize our main findings:

\begin{enumerate} 

\item{} Our new images confirm the existence of complex filamentary structures in the Toothbrush up to 8\,GHz. We detect the other two fainter relics and the radio halo up to 4\,GHz. \\

\item{} The radio integrated emission from the entire Toothbrush and subregions (B1, B2, and B3) closely follow a power-law, with an integrated spectral index of  $\alpha=-1.16\pm0.02$. We do not find any evidence of spectral steepening of the relic emission at any frequency below 8\,GHz. \\

\item{} The Toothbrush subregions B1, B2 and B3 show nearly identical spectra across a $\sim 2 \, \rm Mpc$ scale, implying a very similar combination of shock properties across the shock front or the robust convergence of the spectral shape when averaged over some distribution of underlying properties. \\

\item{} Comparison with a recent numerical simulation of radio relics suggests that such a narrow variation in the integrated spectra are plausible when projection and cooling effects of electrons are properly taken into account. The simulation shows that there is a broad distribution of Mach number at the shock front, and that the ones at the high end dominate the spectral distribution. It seems plausible that the combination of the tail in the distribution towards high Mach numbers and the dependence of the radio luminosity on the Mach number determines the spectral index of the radio emission.\\

\item{} The spectral shapes inferred from spatially resolved regions reveal less homogeneity because the spectra show curvature. However, the observed steepening in these individual regions is still less than expected from any standard aging models. The trajectories in the color-color diagram matches reasonably well with the KGJP-model indicating that the spectral index and curvature found for the Toothbrush can be understood assuming a propagating shock front, which is slightly inclined to the line of sight.\\

\item{} The integrated spectra of portions of relics E and D show spectral shapes which are close to power laws, with different slopes at different locations. Other regions show some steepening at high frequencies.  The types of inhomogeneities and/or physical conditions that would lead to this type of behavior are not yet clear.\\

\item{} In a simplistic model, we find that the downstream material emits an energy equivalent to about 1\,\%  of the thermal energy due to synchrotron losses. This implies that the mechanism responsible for accelerating electrons to relativistic energies must be very efficient and that magnetic fields in the emitting volume below $\sim 1\, \rm \upmu G$ are unlikely. \\

\item{}The integrated spectral index of the radio halo follows a well defined power law up to 3\,GHz. A color-color analysis of regions within the radio halo shows no sign of spectral curvature.\\

\item{} Spectral analysis of the relic+halo region provides evidence that the two regions overlap along the line of sight. We present a simple model in which the observed positive curvature in the relic+halo spectra is well explained by the simple superposition, of an aging population of radio electrons injected by the relic shock, and of a different (physically disconnected) population of electrons in the radio halo. 

\end{enumerate}

\section*{Acknowledgements}
We thank the anonymous reviewer for the constructive feedback. K.R., F.V., and D.W. acknowledges financial support from the ERC  Starting Grant "MAGCOW", no. 714196.  Part of this work was performed at Th\"uringer Landessternwarte Tautenburg, Germany. The cosmological simulations in this work were performed using the ENZO code (http://enzo-project.org). The authors gratefully acknowledge the Gauss Centre for Supercomputing e.V. (www.gauss-centre.eu) for supporting this project by providing computing time through the John von Neumann Institute for Computing (NIC) on the GCS Supercomputer JUWELS at J\"ulich Supercomputing Centre (JSC), under projects no. HHH42 and {\it stressicm} (PI F.Vazza) as well as HHH44 (PI D. Wittor).
Partial support for the work of L.R. is provided by U.S. National Science Foundation grant AST17-14205 to the University of Minnesota. A.D. acknowledges support by the BMBF Verbundforschung under the grant 05A17STA. RK acknowledges the support of the Department of Atomic Energy, Government
of India, under project no. 12-R\&D-TFR-5.02-0700 and support from the
DST-INSPIRE Faculty Award of the Government of India. We also acknowledge the usage of online storage tools kindly provided by the INAF Astronomical Archive (IA2) initiave (http://www.ia2.inaf.it).

The National Radio Astronomy Observatory is a facility of the National Science Foundation operated under cooperative agreement by Associated Universities. We thank the staff of the GMRT that made these observations possible. GMRT is run by the National Centre for Radio Astrophysics of the Tata Institute of Fundamental Research. This paper is based (in part) on data obtained with the International LOFAR Telescope (ILT). LOFAR (van Haarlem et al. 2013) is the Low Frequency Array designed and constructed by ASTRON. It has facilities in several countries that are owned by various parties (each with their own funding sources), and that are collectively operated by the ILT foundation under a joint scientific policy.

\bibliographystyle{aa}

\bibliography{rxs42.bib}

\end{document}